\documentclass[authoryear,preprint,review,12pt]{elsarticle}

\usepackage[colorlinks=true, urlcolor=citecolor, linkcolor=citecolor,
citecolor=citecolor]{hyperref}
\usepackage{amsmath,enumerate,multirow,verbatim,
ragged2e,array,multirow,amssymb}
\usepackage[figuresright]{rotating}
\usepackage[margin=1.0in]{geometry}

\def\bSig\mathbf{\Sigma}

\usepackage{lineno,hyperref}
\modulolinenumbers[5]

\journal{Computational Statistics \& Data Analysis}

\begin{document}

\begin{frontmatter}

\title{Identification of relevant subtypes via preweighted sparse clustering}

\author[har]{Sheila Gaynor}

\author[unc]{Eric Bair\corref{cor1}\fnref{unc2}}
\ead{ebair@email.unc.edu}

\address[har]{Department of Biostatistics, Harvard University, Boston, MA, U.S.A. }
\address[unc]{Departments of Endodontics and Biostatistics,
  University of North Carolina at Chapel Hill, Chapel Hill,
  North Carolina, U.S.A.}
\fntext[unc2]{Contact Information: School of Dentistry, CB \#7450,
  Chapel Hill, NC 27599-7450, 919-537-3276}
\cortext[cor1]{To whom correspondence should be addressed.}
\fntext[boo]{Supplementary material (including the source code used to generate the
tables, Figures \ref{F:motiv_feature_wts} and \ref{F:psc_oppera_ws},
and the leukemia microarray data set) is available with this paper at
the journal website.}

\begin{abstract}
Cluster analysis methods are used to identify homogeneous subgroups in
a data set. In biomedical applications, one frequently applies cluster
analysis in order to identify biologically interesting subgroups. In
particular, one may wish to identify subgroups that are associated
with a particular outcome of interest. Conventional clustering methods
generally do not identify such subgroups, particularly when there are
a large number of high-variance features in the data set. Conventional
methods may identify clusters associated with these high-variance
features when one wishes to obtain secondary clusters that are more
interesting biologically or more strongly associated with a particular
outcome of interest. A modification of sparse clustering can be used
to identify such secondary clusters or clusters associated with an
outcome of interest. This method correctly identifies such
clusters of interest in several simulation scenarios. The method is
also applied to a large prospective cohort study of temporomandibular
disorders and a leukemia microarray data set.
\end{abstract}


\end{frontmatter}


\section{Introduction}
\label{S:intro}
In biomedical applications, cluster analysis is frequently used to
identify homogeneous subgroups in a data set that provide information
about a biological process of interest. For example, in microarray
studies of cancer, a common objective is to identify cancer subtypes
that are predictive of the prognosis (survival time) of cancer
patients \citep{aB01, tS01, vV02, aR02, jL04, lB04}. In studies of
chronic pain conditions, such as fibromyalgia or temporomandibular
disorders (TMD), one may wish to develop a more precise case
definition for the condition of interest by identifying subgroups of
patients with similar clinical characteristics \citep{rJ88, sB02a,
pD03, bH05}. However, conventional clustering methods (such as
k-means clustering and hierarchical clustering) may produce
unsatisfactory results when applied to these types of problems.

Identification of relevant clusters in complex data sets
presents several challenges. It is common that only a subset of the
features will have different means with respect to the clusters.
This is particularly true in genetic studies, where the
majority of the genes are not associated with the outcome of
interest. Moreover, it is possible that some other subset of the
features form clusters that are not associated with the outcome of
interest. In genetic studies, given that genes work in pathways, genes in
the same pathway are likely to form clusters even if the pathway is
not associated with the biological outcome of interest.

\begin{figure}[h!]
  \caption{Artificial data set illustrating the limitations of
    conventional clustering methods. Suppose observations 1-100 are
    controls and observations 101-200 are cases. In this situation,
    one would be interested in the clusters formed by features 1-50,
    but most existing clustering methods would identify the clusters
    formed by features 51-250 (and hence does not identify the cluster
    formed by features 1-50).}
  \label{F:schem1}
  \centering\includegraphics[height=14cm,angle=90]{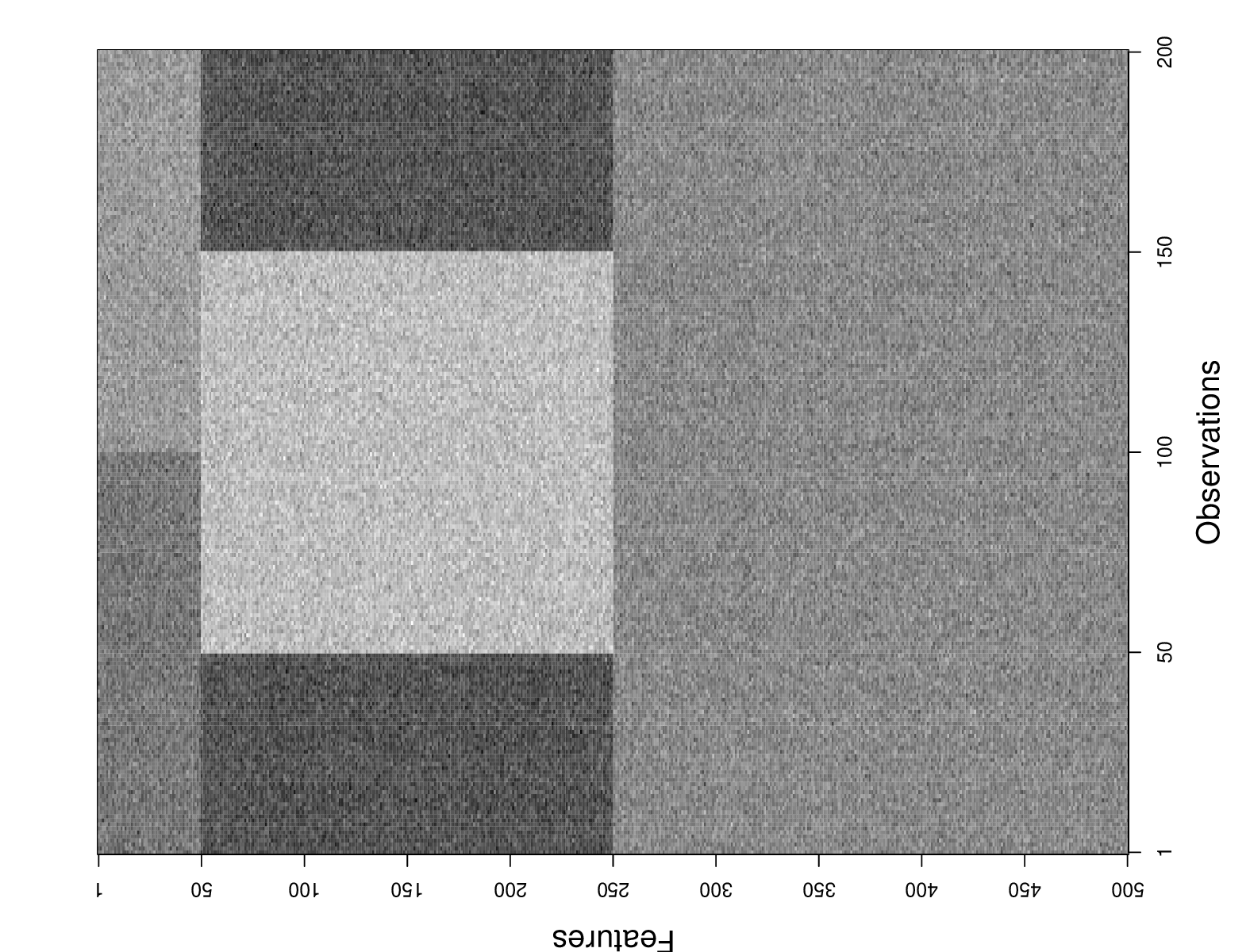}
\end{figure}

As a motivating example, consider the artificial data set
represented in Figure \ref{F:schem1}. This depicts a standard
clustering scenario in which we are seeking to cluster $n$
observations based on $p$ measured features,
where the data is in the form of a $p \times n$ matrix.
We see that two sets of clusters exist in this data: a set of clusters
where the observations differ with respect to features 1-50, and a
separate set of clusters where observations differ with respect to
features 51-250. Also, note that the difference between the cluster
means is much greater for the clusters formed by features 51-250 than
it is for the clusters formed by features 1-50. Thus, when
conventional clustering methods are applied to this data set, they
will most likely identify the clusters corresponding to features
51-250. We define these clusters to be primary clusters, as they are
most likely to be identified by conventional clustering
methods. However, if observations 1-100 are controls and observations
101-200 are cases, then we would be interested in the clusters
corresponding to features 1-50, which would not be identified by most
existing clustering methods. Such a cluster that differs with respect
to different features than the primary clusters (and hence will not be
identified by most conventional clusters methods that identify the
primary clusters) will be referred to as a secondary cluster. See
\citet{NT08} for a more detailed discussion of this problem.

Note that labeling one set of clusters as ``primary'' and other sets
of clusters as ``secondary'' or ``tertiary'' is done for convenience
and that these labels are somewhat arbitrary. For a given data set,
one may obtain different ``primary clusters'' depending on the
clustering method and distance metric used (among other possible
factors). Also, there may be some overlap between the features that
differ with respect to the primary clusters and the features that
differ with respect to the secondary clusters. However, the list of
features the differ with respect to the primary and secondary clusters
should not be identical. (If they differed with respect to exactly the
same set of features, one could identify the ``secondary cluster'' by
simply increasing the number of clusters $k$. Specialized methods are
only needed when there are features that differ with respect to the
secondary clusters but not the primary clusters.)

A number of methods exist for clustering data sets when the clusters
differ with respect to only a subset of the features \citep{GC02,
FM04, BT04, RD06, PS07, dK10, WT10}. In particular, the method of
\citet{NT08} is designed specifically for the situation described in
Figure \ref{F:schem1}. However, many of these methods are
computationally intensive, and their running times may be
prohibitive when applied to high-dimensional data sets. More
importantly, with the exception of the method of \citet{NT08}, these
methods only produce a single set of clusters. If the clusters
identified by the method are not related to the biological outcome of
interest, there is no simple way to identify the more relevant
secondary clusters. Also, these methods generally do not consider an
outcome variable or any other biological information that could help
identify the clusters of interest. In other words, if these methods
are applied to a data set similar to Figure \ref{F:schem1}, they are
likely to produce clusters that are not related to the outcome of interest.

\begin{figure}[h!]
  \caption{Artificial example of a situation where the outcome
    variable is a ``noisy surrogate'' for the true clusters. In this
    artificial example, the density functions of the outcome variable
    for observations in each of two clusters are shown above.
    Observations in cluster 2 are more likely to have higher values of
    the outcome variable than observations in cluster 1, but there is
    considerable overlap between the two groups. Thus, classifying
    observations to clusters based solely on the outcome variable will
    result in a high misclassification error rate.}
  \label{F:schem2}
  \centering\includegraphics[height=10cm]{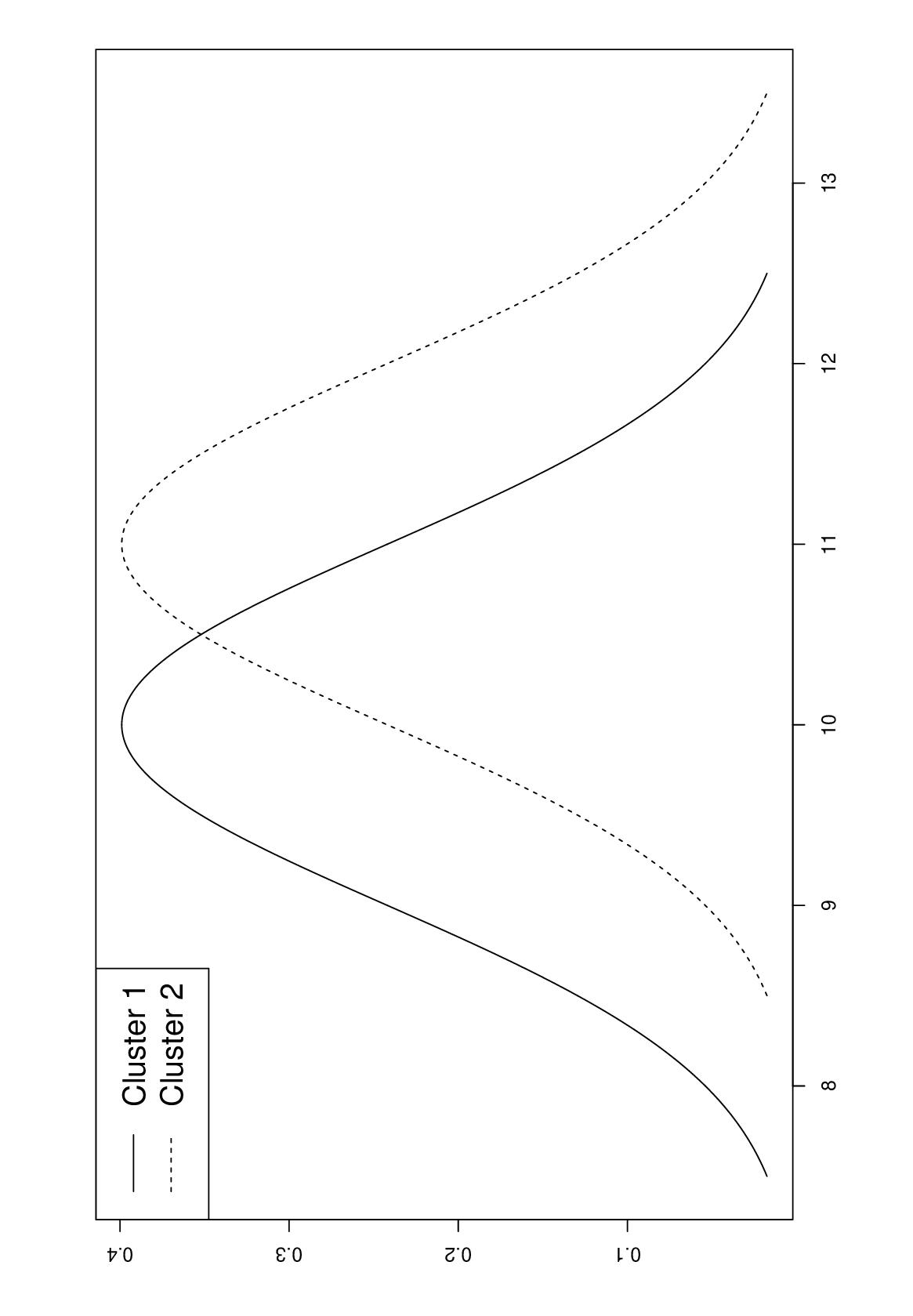}
\end{figure}

The problem of identifying clusters associated with an outcome
variable has also not been studied extensively \citep{eB13}. In many
situations, there is an outcome variable that is a  ``noisy
surrogate'' \citep{BT04, eB06} for the true clusters. For example, in
genetic studies of cancer, it is believed that there are underlying
subtypes of cancer with different genetic aberrations, and some
subtypes may be more responsive to treatment \citep{aR02, lB04,
BT04}. These subtypes cannot be observed directly, but a surrogate
variable such as the patient's survival time may be available. In
other words, the outcome variable provides some information about the
clusters of interest, but the true cluster assignments are still
unknown for all observations. An artificial example of this situation
is shown in Figure \ref{F:schem2}. In this example, the mean of the
outcome variable for observations in cluster 2 is higher than the mean
of the outcome variable for observations in cluster 1. However, there
is considerable overlap in the distributions. Thus, higher values of
the outcome variable increase the likelihood that an observation
belongs to cluster 2, but any classifier that attempts to predict the
cluster based on the outcome variable will have a high error rate.

We propose a novel clustering method that is applicable in situations
where one wishes to identify secondary clusters associated with an
outcome of interest, such as the scenario illustrated in Figure
\ref{F:schem1}. It is based on a modification of the ``sparse
clustering'' algorithm of \citet{WT10}, which we call preweighted
sparse clustering. It can be applied both to the general problem of
identifying secondary clusters in data sets and to the special case
where one wishes to identify clusters associated with an outcome
variable. We will show that our proposed method produces more accurate
results than competing methods in several simulated data sets and
apply it to real-world studies of chronic pain and cancer.

\section{Methods}
\label{S:methods}
This section will begin by briefly describing several existing methods
for identifying clusters associated with a biological process of
interest. We will then describe our proposed method as well as the
simulated and real data sets to which the proposed method will
be applied.

\subsection{Related Clustering Methods}
\subsubsection{Sparse Clustering} \label{S:sparse_clust}
Suppose that we wish to cluster the $p \times n$ data matrix $X$,
where $p$ is the number of features and $n$ is the number of observations.
 Assume that the clusters only differ with respect to some
subset of the features. \citet{WT10} propose a method called
``sparse clustering'' to solve this problem. A brief description of
the sparse clustering method is as follows: Let $d_{i,j,j'}$ be any
dissimilarity measure between observations $j$ and $j'$ with respect
to feature $i$. (Throughout the remainder of this discussion, we will
assume that $d_{i,j,j'} = (X_{ij} - X_{ij'})^2$ the Euclidean distance
between $X_{ij}$ and $X_{ij'}$.) Then \citet{WT10} propose to identify
clusters $C_1, C_2, \ldots, C_K$ and weights $w_1, w_2, \ldots, w_p$
that maximize the weighted between-cluster sum of squares
\begin{equation} \label{E:SClustCrit}
\sum_{i=1}^p w_i \left( \frac{1}{n} \sum_{j=1}^n \sum_{j'=1}^n
  d_{i,j,j'} - \sum_{k=1}^K \frac{1}{n_k} \sum_{j,j' \in C_k}
  d_{i,j,j'} \right)\text{,}
\end{equation}
subject to the constraints $\sum_i w_i^2=1$, $\sum_i |w_i|<s$, and
$w_i \geq 0$ for all $i$, where $s$ is a tuning parameter and $n_k$ is
the number of elements in cluster $k$. Note that the $\sum_i |w_i|<s$
constraint forces some of the weights to 0 for sufficiently small
values of $s$, resulting in clusters that are based on only a subset
of the features (hence the term ``sparse clustering''). This is
similar to the constraint used in lasso regression \citep{rT96} to
produce a sparse set of predictor variables. To maximize
(\ref{E:SClustCrit}), \citet{WT10} use the following algorithm:
\begin{enumerate}
  \item Initialize the weights as $w_1=w_2= \cdots = w_p =
    1/\sqrt{p}$.
  \item Fix the $w_i$'s and identify $C_1, C_2, \ldots, C_K$ to
    maximize (\ref{E:SClustCrit}). This can be done by applying the
    standard k-means clustering method to the $n \times n$
    dissimilarity matrix where the $(j,j')$ element is $\sum_i w_i
    d_{i,j,j'}$. \label{en:max_C}
  \item Fix the $C_i$'s and identify $w_1, w_2, \ldots, w_p$ to
    maximize (\ref{E:SClustCrit}) subject to the constraints that
    $\sum_i w_i^2=1$ and $\sum_i |w_i|<s$. See \citet{WT10} for a
    description of how the optimal $w_i$'s are
    calculated. \label{en:max_w} 
  \item Repeat steps \ref{en:max_C} and \ref{en:max_w} until
    convergence.
\end{enumerate}
This procedure requires a user to choose the number of clusters $k$
and the tuning parameter $s$. We will not discuss methods for choosing
these parameters; see \citet{WT10} for an algorithm for choosing $s$,
and see \citet{TWH01}, \citet{SJ03}, or \citet{TW05} for several
possible methods for choosing $k$.

Although this method correctly identifies clusters of interest in many
situations, it tends to identify clusters that are dominated by highly
correlated features with high variance, which may not be interesting
biologically. It also does not consider the values of any outcome
variables that may exist. Thus, in the situation illustrated in Figure
\ref{F:schem1}, there is no guarantee that the clusters identified by
this method will be associated with the outcome of interest.

\subsubsection{Complementary Clustering}
Methods have been developed to identify secondary clusters of interest
that may be obscured by ``primary'' clusters consisting of large
numbers of high variance features (such as the situation illustrated
in Figure \ref{F:schem1}).  \citet{NT08} proposed a method for
uncovering such clusters, called complementary hierarchical
clustering. Again assume that we wish to cluster the $p \times n$ data
matrix $X$. The first step of this method performs traditional
hierarchical clustering on $X$. This set of hierarchical clusters is
used to generate a new matrix $X'$ that is defined to be the expected
value of the residuals when each row of $X$ is regressed on the group
labels when the hierarchical clustering tree is cut at a given
height. The expected value is taken over all possible cuts. This has
the effect of removing high variance features that may be obscuring
secondary clusters. Traditional hierarchical clustering is then
performed on this modified matrix $X'$, yielding secondary
clusters. \citet{WT10} proposed a modification of this procedure
(called ``sparse complementary clustering'') using a variant of the
methodology described in Section \ref{S:sparse_clust}.

One significant limitation of these methods is the fact that they are
only applicable to hierarchical clustering. To our knowledge there are
currently no published methods for identifying secondary clusters
based on partitional clustering methods such as k-means clustering.

\subsubsection{Semi-Supervised Clustering Methods}
The situation where the observed outcome variable is a noisy
surrogate variable for underlying clusters is very common in
real-world problems. However, there are relatively few clustering
methods that are applicable for this type of problem \citep{eB13}.
\citet{BT04} proposed a method that they called ``supervised
clustering.'' Supervised clustering performs conventional k-means
clustering or hierarchical clustering using only a subset of the
features. The features are selected by identifying a fixed number of
features that have the strongest univariate association with the outcome
variable. For example, if the outcome is dichotomous, one would
calculate a t-statistic for each feature to test the null hypothesis
of no association between the feature and the outcome and then perform
clustering using only the features with the largest (absolute)
t-statistics. \citet{dK10} proposed a method called ``semi-supervised
recursively partitioned mixture models'' (or ``semi-supervised
RPMM''). This method is similar to the supervised clustering method of
\citet{BT04} in that one first calculates a score for each feature
(such a t-statistic) that measures the association between that
feature and the outcome and then performs clustering using only the
features with the largest univariate scores. The difference between
semi-supervised RPMM and supervised clustering is that semi-supervised
RPMM applies the RPMM algorithm of \citet{aH08} to the surviving
features rather than a more conventional k-means or hierarchical
clustering model.

These methods have successfully identified clinically relevant
subtypes of cancer in many different studies \citep{BT04, lB04, aC08,
dK10}. However, these methods have significant limitations. In
particular, both supervised clustering and semi-supervised RPMM
require a user to choose the number of features that are used to form
the clusters, and the results of these methods can depend heavily on
the number of ``significant'' features selected. Moreover, it is very
unlikely that these methods will successfully identify the truly
significant features that define the clusters while excluding
irrelevant features.

\subsection{Preweighted Sparse Clustering} \label{S:psc}
To overcome the limitations of these methods, we propose the following
modification of sparse clustering, which we call preweighted sparse
clustering. The preweighted sparse clustering algorithm is described
below:
\begin{enumerate}
  \item Run the sparse clustering algorithm, as described previously.
  \item For each feature, calculate the F-statistic, $F_i$, (and
    associated p-value $q_i$) for testing the null hypothesis that the
    mean value of the feature $i$ does not vary across the clusters.
  \item For each feature $i$, define:
    \[
    w_i =
    \begin{cases}
      1/\sqrt{m} &\text{if $q_i \geq \alpha$} \\
      0 &\text{otherwise}
    \end{cases}\text{,}
    \]
    where $m$ is the number of $q_i$'s such $q_i \geq \alpha$.
  \item Run the sparse clustering algorithm using these $w_i$'s
    (beginning with step \ref{en:max_C}) and continuing until
    convergence.
\end{enumerate}
In other words, the preweighted sparse clustering algorithm first
performs conventional sparse clustering. It then identifies features
whose mean values differ across the clusters. Then the sparse
clustering algorithm is run a second time, but rather than giving
equal weights to all features as in the first step, this preweighted
version of sparse clustering assigns a weight of 0 to all features
that differed across the first set of clusters. The motivation is that
this procedure will identify secondary clusters that would otherwise
be obscured by clusters that have a larger dissimilarity measure (such
as the situation illustrated in Figure \ref{F:schem1}).

This procedure requires one to choose a p-value threshold $\alpha$ for
deciding which features should be given nonzero weight. An obvious
choice is $\alpha=0.05/p$, where $p$ is the number of
features. However, the user may choose a less or more  stringent
cutoff depending on the sample size and other considerations. (Note
that the F-statistic was calculated after clustering was performed, so
the test statistic need not have an F distribution under the null
hypothesis of no mean difference between clusters. Thus, these
p-values should not be used to conclude that a given feature is
associated with the clusters; this procedure is used only as a
filtering technique. Indeed, the problem of identifying the features
that differ with respect to clusters is a difficult problem that is
beyond the scope of the present study.) Also note that this procedure
may be repeated multiple times if one wishes to identify tertiary or
higher order clusters.

If desired, one may normalize the data such that all features have
mean 0 and standard deviation 1 before applying the methodology. This
is recommended for most applications to avoid giving undue weight to
features with higher variance. Unless otherwise noted, the data will
be normalized before applying preweighted sparse clustering in all
subsequent examples.

\subsection{Supervised Sparse Clustering}
The preweighted sparse clustering algorithm described above is an
unsupervised method, since it does not require or use an outcome
variable. If an outcome variable is available and the objective is to
identify clusters associated with the outcome variable, one may use
the following variant of preweighted sparse clustering to incorporate
such data, which we call supervised sparse clustering.  The supervised
sparse clustering procedure is described below:
\begin{enumerate}
\item Let $T_i$ be a measure of the strength of the association
  between the $i$th feature and the outcome variable. (If the outcome
  variable is dichotomous, $T_i$ could be a t-statistic, or if the
  outcome variable is a survival time, $T_i$ could be a univariate Cox
  score.) Let $T_{(1)}, T_{(2)}, \ldots, T_{(p)}$ denote the order
  statistics of the $T_i$'s.
\item Run the sparse clustering algorithm with initial
  weights $w_1, w_2, \ldots, w_p$, where
  \[
  w_j =
  \begin{cases}
    1/\sqrt{m} &\text{if $|T_i| \geq |T_{(p-m+1)}|$} \\
    0 &\text{otherwise}
  \end{cases}\text{.}
  \]
\item Run the standard sparse clustering algorithm using these $w_i$'s
    (beginning with step \ref{en:max_C}) and continuing until
    convergence.
\end{enumerate}
In other words, supervised sparse clustering chooses the initial
weights for the sparse clustering algorithm by giving nonzero weights
to the features that are most strongly associated with the outcome
variable. Note that that no initial clustering step is required. This
is similar to the semi-supervised clustering method of \citet{BT04}
and the semi-supervised RPMM method of \citet{dK10}.

The supervised sparse clustering procedure requires the choice of a
tuning parameter $m$, which is the number of features to be given
nonzero weight in the first step. Our experience suggests that the
procedure tends to give very similar results for a wide variety of
different values of $m$; therefore, optimizing the procedure with respect to
this tuning parameter is unnecessary. As a default we suggest
$m=\sqrt{p}$, where $p$ is the number of features. We will use this
default throughout this manuscript unless otherwise noted.

\subsection{Simulated Data Sets} \label{S:m_sim}
A series of simulations were performed to evaluate the performance of
our proposed methods and to compare them to the results of existing
methods. Several additional simulation studies are described in
Section \ref{S:supp_sims} in the Supplementary Materials.
\subsubsection{A Motivating Example} \label{S:motiv}
A single $50 \times 100$ data set was generated as follows:
\begin{equation} \label{E:motiv}
X_{ij} =
\begin{cases}
  6 + \epsilon_{ij} &\text{if $i \leq 10, j \leq 50$} \\
  -6 + \epsilon_{ij} &\text{if $i \leq 10, j > 50$} \\
  9 + \epsilon_{ij} &\text{if $11 \leq i \leq 20$, $j \leq 50$, and
    $j$ odd} \\
  3 + \epsilon_{ij} &\text{if $11 \leq i \leq 20$, $j \leq 50$, and
    $j$ even} \\
  -3 + \epsilon_{ij} &\text{if $11 \leq i \leq 20$, $j \leq 50$, and
    $j$ odd} \\
  -9 + \epsilon_{ij} &\text{if $11 \leq i \leq 20$, $j \leq 50$, and
    $j$ even} \\
  3 + \epsilon_{ij} &\text{if $i \leq 21 \leq 30$, $j$ odd} \\
  -3 + \epsilon_{ij} &\text{if $i \leq 21 \leq 30$, $j$ even} \\
  \epsilon_{ij} &\text{if $i > 30$}\text{.}
\end{cases}
\end{equation}
In this data set, the primary clusters are defined by rows 1-20 and
the secondary clusters are defined by rows 11-30. The objective of
this simulation is to determine if preweighted sparse clustering can
identify both the primary and secondary clusters and assign nonzero
weight to the appropriate features.
\subsubsection{Preweighted Sparse Clustering} \label{S:m_pws_sim}
\begin{figure}[h!]
  \caption{Schematic illustration of the simulated data set. The top
    $p_e$ rows correspond to the ``primary clusters'' (i.e., ``Effect
    1'') and the bottom $p_e$ rows correspond to the ``secondary
    clusters'' (i.e., ``Effect 2''). Reprinted with permission from
    \citet{NT08}.}
  \label{F:simschem}
  \centering\includegraphics[height=7cm]{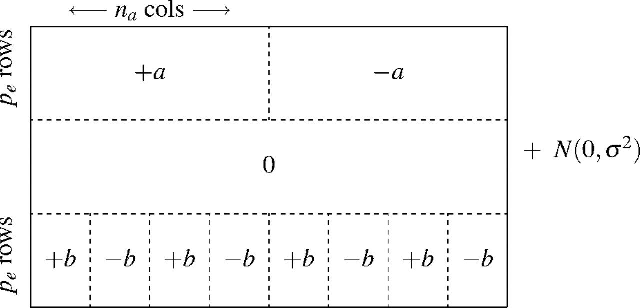}
\end{figure}

We generated a series of simulated data sets to evaluate the
performance of preweighted sparse clustering and compare it to the
complementary hierarchical clustering method of \citet{NT08} and the complementary
hierarchical sparse clustering method of \citet{WT10}. We generated simulated
data sets similar to the simulated data sets in \citet{NT08}, who
generated a series of $p \times 12$ data matrices as follows:
\begin{equation} \label{E:Nowak_sims}
X_{ij} =
\begin{cases}
  a + \epsilon_{ij} &\text{if $i \leq p_{e}, j \leq n_a$} \\
  -a + \epsilon_{ij} &\text{if $i \leq p_{e}, j > n_a$} \\
  \epsilon_{ij} &\text{if $p_{e} < i < p - p_{e} $} \\
  b + \epsilon_{ij} &\text{if $i \geq p - p_{e},\, j$ odd} \\
  -b + \epsilon_{ij} &\text{if $i \geq p - p_{e},\, j$ even}
\end{cases}\text{.}
\end{equation}
Here, the $\epsilon_{ij}$'s are iid normal random variables with mean
0 and standard deviation $\sigma$. See Figure \ref{F:simschem} for a
graphical illustration of this data set. The expectation is that the
first $p_e$ rows (i.e., ``Effect 1'') will be identified as the
``primary clusters'' and the final $p_e$ rows (i.e., ``Effect 2'')
will be identified as the ``secondary clusters'' .

We considered four simulation scenarios (similar to the four
simulation scenarios considered in \citet{NT08}). We let $a=6$ in all
four scenarios. Unless otherwise specified, we also let $b=3$,
$\sigma=1$, and $n_a=6$ for each simulation scenario. For the first
three scenarios, 1000 matrices were generated with $p=50$ and
$p_e=20$. In the first scenario, we varied the value of $b$. In the
second scenario, we varied the value of $\sigma$, and in the third
scenario, we varied $n_a$. In the final scenario, we generated 100
matrices with $p=2000$ and varied the value of $p_e$. (The first three
scenarios are identical to the simulations of \citet{NT08}; the final
scenario was modified slightly for computational reasons.)

Preweighted sparse clustering, complementary hierarchical clustering,
and complementary hierarchical sparse clustering were applied to each
simulated data set. The number of clusters was fixed to be $k=2$ for
all methods. Both hierarchical clustering methods used average
linkage. For complementary hierarchical clustering, two distance
metrics were considered. The first metric defined the distance
between observations $j$ and $j'$ was defined to be
$1-\mbox{corr}(x_{\cdot j}, x_{\cdot j'})$, which is the same distance
metric used by \citet{NT08}. The second distance metric was Euclidean
distance. Complementary hierarchical sparse clustering only used
Euclidean distance, since the correlation distance metric is not
implemented for that procedure.

Each set of clusters identified by each method was compared to the
true cluster labels for both Effect 1 and Effect 2. The number of
simulations for which each method correctly identified Effect 1 and/or
Effect 2 was recorded. The results of a clustering procedure were
considered to be incorrect if one or more observations were assigned
to the incorrect cluster.

\subsubsection{Supervised Sparse Clustering}
In an additional set of simulations, we  generated a series of 1000 simulated data sets
to test the supervised sparse clustering algorithm. Specifically, we generated
1000 $5000 \times 200$ data matrices $X$ where
\[
X_{ij} =
\begin{cases}
  1 + \epsilon_{ij} &\text{if $i \leq 50, j \leq 100$} \\
  2 + \epsilon_{ij} &\text{if $i \leq 50, j > 100$} \\
  2I(u_{ij}<0.4) + \epsilon_{ij} &\text{if $51 \leq i \leq 100$} \\
  0.5I(u_{ij}<0.7) + \epsilon_{ij} &\text{if $101 \leq i \leq 200$} \\
  1.5I(u_{ij}<0.3) + \epsilon_{ij} &\text{if $201 \leq i \leq 300$} \\
  \epsilon_{ij} &\text{if $i>300$}
\end{cases}\text{.}
\]
Here $I(x)$ is an indicator function, and the $u_{ij}$'s are iid
uniform random variables on $(0,1)$. The $\epsilon_{ij}$'s are iid
standard normal, as before. We also defined the binary outcome
variable $y$ as follows:
\[
y_i =
\begin{cases}
  0 + I(u_i < 0.3) &\text{if $1 \leq i \leq 100$} \\
  1 - I(u_i < 0.3) &\text{if $101 \leq i \leq 200$}
\end{cases}\text{.}
\]
(In the above, once again $I(x)$ is an indicator function and the
$u_i$'s are iid uniform random variables on $(0,1)$.) This simulation
is similar to the scenario illustrated in Figure \ref{F:schem1}. We
assume that the first 50 features are the biologically relevant
features of interest. In other words, a clustering algorithm that
achieves perfect accuracy should assign observations 1-100 to one
cluster and observations 101-200 to a separate cluster. Features
51-100, 101-200, and 201-300 also form clusters, but these clusters
are not related to the biological outcome of interest. The outcome
variable $y$ also observed is a ``noisy surrogate'' for the
true clusters. This $y$ is related to the true clusters, but 30\% of
the $y_i$'s are misclassified. This is consistent with what we might
expect to observe in a study of chronic pain, where the only observed
outcome variable is a patient's subjective pain report, which is not
always a reliable indicator of case status.

The objective of this simulation is to determine if supervised sparse
clustering can correctly identify the clusters that are associated
with the $y_i$'s, as opposed to the other sets of clusters that are
not related to the outcome. Supervised sparse clustering was applied
to each of the 1000 simulated data sets. Three other methods were also
considered, namely conventional sparse k-means clustering, the
semi-supervised k-means clustering method of \citet{BT04}, and
conventional k-means clustering on the first three principal
components of the data set. The number of clusters was fixed to be
$k=2$ under all methods. We also attempted to apply the
semi-supervised RPMM method of \citet{dK10} to these simulated
data sets, but in each case the procedure returned a singleton
cluster.  The number of observations assigned to the incorrect cluster
was recorded for each method for each simulation.

\subsection{OPPERA Data}
Orofacial Pain: Prospective Evaluation and Risk Assessment (OPPERA) is
a prospective cohort study designed to identify risk factors for
temporomandibular disorders (TMD). OPPERA recruited a total of $3258$
TMD-free study subjects at four U.S. study sites from May 2006 to
November 2008. Numerous putative risk factors for first-onset TMD were
evaluated at the time of enrollment, and after enrollment each
participant completed a quarterly follow up questionnaire assessing
TMD pain symptoms. Those reporting symptoms were invited for a follow
up exam to determine if they had developed first-onset TMD. The median
follow up period was $2.8$ years, and a total of $260$ participants
developed TMD over the course of the study. For a more detailed
description of the OPPERA study, see \citet{wM11}, \citet{gS11} or
\citet{eB13b}.

We applied our clustering algorithms to the baseline data collected in
OPPERA. Specifically, we included all of the measures of experimental
pain sensitivity, psychological distress, and autonomic function. See
\citet{jG11}, \citet{rF11}, and \citet{wM11b} for a more detailed
description of these variables. A total of $116$ predictor variables
were used, including $33$ measures of experimental pain sensitivity,
$39$ measures of psychological distress, and $44$ measures of
autonomic function. The primary outcome of interest is time until the
development of first-onset TMD. Since some participants did not
develop first-onset TMD before the end of the follow up period, the
outcome was treated as a censored survival time.

In our analysis of the OPPERA data, we applied the preweighted sparse
clustering algorithm as outlined in Section \ref{S:psc}. Conventional
sparse clustering was applied to the data set (also with $k=2$), after which the
features that showed strongest mean differences across the clusters
were given a weight of 0 when the preweighted version of sparse
clustering was applied. The preweighted version was then applied for a
second time in the same manner to identify tertiary clusters. All
features were normalized to have mean 0 and standard deviation 1 prior
to performing the clustering. The association between both the primary
clusters and secondary clusters and the time until first-onset TMD was
evaluated using Cox proportional hazards models. Complementary
hierarchical clustering was also applied to both data sets for
comparison. Complementary sparse hierarchical clustering was not
considered for computational reasons.

We also applied our supervised sparse clustering, sparse clustering,
preweighted sparse clustering, semi-supervised clustering, and
clustering on the (first five) principal component scores to the
OPPERA data. We let $k=2$ for each method. To verify that associations
between clusters and first-onset TMD are not the results of
overfitting, the data set was randomly partitioned into a training set
and a test set with an equal number of cases of first-onset TMD in
both partitions. To identify the ``most significant'' predictors of
first-onset TMD before applying supervised sparse clustering and
semi-supervised clustering, the association between each feature and
first-onset TMD was evaluated by calculating the univariate Cox score
for each feature. See \citet{dB02} or \citet{BT04} for more
information. Each clustering method was applied to the training data
and a lasso model \citep{rT96, FHT10} was fit to the training data to
predict the resulting clusters. This lasso model was then used to
predict the clusters on the test data. The association between the
predicted clusters and first-onset TMD was evaluated by fitting a Cox
proportional hazards model on the test data. Note that the test data
was not used to identify the features associated with first-onset TMD
nor to identify the initial clusters, so any association between the
predicted clusters on the test data and first-onset TMD cannot be
explained by overfitting.

\subsection{Leukemia Microarray Data}
We applied our supervised sparse clustering algorithm to the leukemia
microarray data of \citet{lB04}. This data set includes data
for 116 subjects with acute myeloid leukemia. Gene expression data for
$6283$ genes are recorded for each subject, as well as survival times
and outcomes. Survival times ranged from $0$ to $1625$ days, with an
average time of $407.1$ days. The objective was to identify genetic
subtypes (i.e. clusters) using the gene expression data that could be
used to predict the prognosis of leukemia patients.

We applied our supervised sparse clustering method to this
data set as well conventional sparse clustering, preweighted sparse
clustering, semi-supervised clustering, and clustering on the PCA
scores. The number of clusters was taken to be 2 in all
methods. Before applying any of the clustering methods, all features
were normalized to have mean 0 and standard deviation 1 and the data
were randomly partitioned into a training set and a test set, each of
which consisted of $58$ observations. Each clustering method was
applied to the training data. To identify the ``most significant''
genes for supervised sparse clustering and semi-supervised clustering,
the association between each gene and survival was evaluated by
calculating the univariate Cox score for each gene. For each set of
clusters, a nearest shrunken centroid model \citep{rT02} was fit to
the clusters in the training data and then applied to the test data to
predict cluster assignments on the test data. (As in the previous
example, clusters were predicted on an independent test set to ensure
that the results are not due to overfitting.) The association between
the predicted clusters in the test set and survival was evaluated
using Cox proportional hazards models for each clustering method.



\section{Results}
\label{S:results}
\subsection{Simulated Data Sets} \label {S:r_sim}

\subsubsection{Motivating Example} \label{S:r_motiv}
\begin{figure}[h!]
  \caption{This figure shows the feature weights for the primary
    clusters as well as the initial and final feature weights for the
    secondary clusters for the motivating example. Note that the
    procedure for identifying secondary clusters initially gives a
    weight of 0 to features 11-20 (since they are also associated with
    the primary clusters) and nonzero weight to features 31-50 (since
    they are not associated with the primary clusters). When the
    procedure terminates, however, features 11-20 have nonzero weight
    and features 30-50 have (effectively) zero weight.}
  \centering\includegraphics[height=13cm]{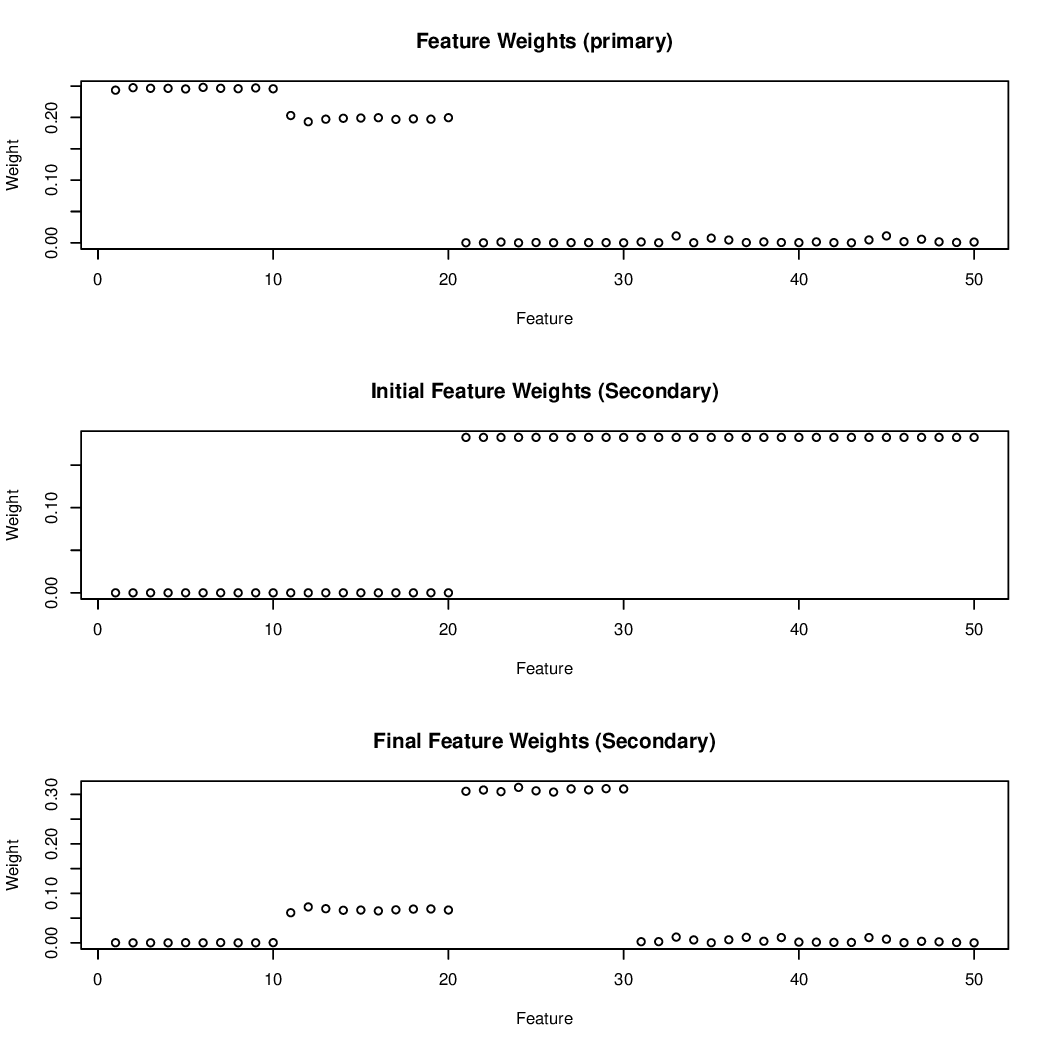}
  \label{F:motiv_feature_wts}
\end{figure}

Preweighted sparse clustering correctly identified both the primary
and secondary clusters in this example. The feature weights for the
primary clusters as well as the initial and final feature weights for
the secondary clusters are shown in Figure
\ref{F:motiv_feature_wts}. The procedure identifies the primary
clusters correctly and assigns nonzero weight to the appropriate
observations. Since preweighted sparse clustering initially assigns
zero weight to features that are associated with the primary clusters
and equal weight to features that are not associated with the primary
clusters, this means that initially features 11-20 (which are
associated with the secondary clusters) have a weight of zero and
features 31-50 (which are not associated with the secondary clusters)
have a nonzero weight. When the procedure terminates, however,
features 11-20 have nonzero weight and features 31-50 have
(effectively) zero weight.

This result is important because it demonstrates that preweighted
sparse clustering and supervised sparse clustering can accurately
identify clusters (and the features that define these clusters) even
if the initial cluster weights give zero weight to some relevant
features and nonzero weight to irrelevant features. Thus, it is not
essential to choose an ``optimal'' set of initial weights since the
procedure tends to correct itself. The implication is that these
methods are robust to giving too many (or too few) features nonzero
weight at the initial step. This is potentially an advantage of
supervised sparse clustering compared to existing supervised
clustering methods \citep{BT04,dK10} that merely cluster on a subset
of the features that are most strongly associated with the
outcome. Once these methods choose a set of features to use for the
clustering, that set of features is fixed, so the results may depend
heavily on the features chosen (and a suboptimal choice of features
may produce poor results). By contrast, supervised sparse clustering
(and preweighted sparse clustering) tends to self-correct so that
relevant features get nonzero weight (even if their initial weight was
zero) and irrelevant features get zero weight (even if their initial
weight was nonzero).

In practice it is often to difficult to determine if a feature weight
produced by sparse clustering is ``significantly'' different from
0. Thus, our procedures do not attempt to find an exhaustive list of
all features associated with the clusters. (One may find a list of at
least some features associated with the clusters by increasing the
value of the tuning parameter $s$, but increasing $s$ too much can
cause features truly associated with the outcome to have zero weight.)
This simulation demonstrates that preweighted sparse clustering tends
to give nonzero weight to the correct features even if there is no
simple way to determine which features have truly nonzero weight.

\subsubsection{Preweighted Sparse Clustering}
The results of the set of simulation scenarios are shown in
Tables \ref{T:Nowak1}, \ref{T:Nowak2}, \ref{T:Nowak3}, and
\ref{T:Nowak4}. Preweighted sparse clustering produced the best
results in the first simulation scenario and generally performed the
best in the second simulation scenario. Specifically, preweighted
sparse clustering correctly identified both the primary
and secondary clusters for small values of $b$ in the first scenario
and for large values of $\sigma$ in the second scenario (although it
appears to be slightly less likely to correctly identify the secondary
clusters in the second scenario for $\sigma \geq 5$). This
indicates that preweighted sparse clustering may produce more accurate
results than the competing methods when the signal to noise ratio is
low (either because the mean difference in the two clusters is small
or the amount of random noise is large). All three methods performed
well in the third simulation scenario, with both complementary
hierarchical methods performing perfectly. The results of the
final simulation were mixed. Complementary hierarchical clustering
never correctly identified both the primary and secondary clusters for
small values of $p_e$ whereas preweighted sparse clustering did
identify both clusters in at least some simulations for all $p_e >
4$. However, complementary hierarchical clustering always correctly
identified at least one of the two clusters, whereas preweighted
sparse clustering sometimes identified neither cluster
correctly. Complementary sparse hierarchical clustering performed the
best on this simulation scenario, with near perfect accuracy even for
small values of $p_e$.

\begin{table}[h!]
  \begin{footnotesize}
  \caption{Results of the first simulation when the values of $b$ (the
    difference between the means of the secondary clusters)
    were varied. The clusters associated with the first $p_e$ rows
    were defined to be ``Effect 1,'' and the clusters associated with
    the final $p_e$ rows were defined to be ``Effect 2.'' ``Effect
    1/Effect 2'' means that the primary clusters identified by the
    procedure correspond to Effect 1 and the secondary clusters
    correspond to Effect 2. ``Effect 1/Neither'' means that the
    primary clusters corresponded to Effect 1 but the secondary
    clusters corresponded to neither Effect 1 nor Effect
    2. PSC=preweighted sparse clustering, CHC=complementary
    hierarchical clustering (using both correlation and Euclidean
    distance), CSHC=complementary sparse hierarchical clustering.}
  \label{T:Nowak1}
{\tabcolsep=4.25pt
  \begin{tabular}{@{}ccccc@{}}
    & b & Effect 1/Effect 2 & Effect 2/Effect 1 & Effect 1/Neither  \\ \hline
    PSC
&0.5	&	419	&	0	&	581	\\
&0.75	&	959	&	0	&	41	\\
&1	&	998	&	0	&	2	\\
&2	&	1000	&	0	&	0	\\
&3	&	997	&	3	&	0	\\
&6	&	512	&	488	&	0	\\ \hline
    CHC (corr.)
&0.5	&	142	&	0	& 	858	\\
&0.75	&	591	&	0	&	409	\\
&1	&	925	&	0	&	75       \\
&2	&	1000	&	0	&	0	\\
&3	&	1000	&	0	&	0	\\
&6	&	488	&	512	&	0	\\ \hline
    CHC (Eucl.)
&0.5	&	282	&	0	& 	718	\\
&0.75	&	826	&	0	&	174	\\
&1	&	987	&	0	&	13       \\
&2	&	1000	&	0	&	0	\\
&3	&	1000	&	0	&	0	\\
&6	&	488	&	512	&	0	\\ \hline
    CSHC
 &0.5	&	0	&	0	&	1000	\\
&0.75	&	0	&	0	&	1000	\\
&1	&	0	&	0	&	1000	\\
&2	&	1000	&	0	&	0	\\
&3	&	1000	&	0	&	0	\\
&6	&	481	&	519	&	0	\\ \hline
  \end{tabular}}
\end{footnotesize}
\end{table}

\begin{table}[h!]
  \begin{footnotesize}
  \caption{Results of the first simulation when the values of
    $\sigma$ (the standard deviation of the simulated data) were
    varied. The clusters associated with the first $p_e$
    rows were defined to be ``Effect 1,'' and the clusters associated
    with the final $p_e$ rows were defined to be ``Effect 2.''
    ``Effect 1/Effect 2'' means that the primary clusters identified
    by the procedure correspond to Effect 1 and the secondary clusters
    correspond to Effect 2. ``Effect 1/Neither'' means that the
    primary clusters corresponded to Effect 1 but the secondary
    clusters corresponded to neither Effect 1 nor Effect
    2. PSC=preweighted sparse clustering, CHC=complementary
    hierarchical clustering (using both correlation and Euclidean
    distance), CSHC=complementary sparse hierarchical clustering.}
  \label{T:Nowak2}
{\tabcolsep=4.25pt
  \begin{tabular}{@{}ccccccc@{}}
    &	$\sigma$	&	Effect 1/Effect 2	&	Effect 2/Effect 1	&	Effect
1/Neither	&	Effect 2/Neither	&	Neither/Neither	\\  \hline
PSC
	&	1	&	994	&	6	&	0	&	0 	&	0	\tabularnewline
	&	2	&	1000	&	0	&	0	&	0	&	0	\tabularnewline
	&	2.5	&	1000	&	0	&	0	&	0	&	0	\tabularnewline
	&	3	&	993	&	0	&	7	&	0	&	0	\tabularnewline
	&	4	&	509	&	0	&	491	&	0	&	0	\tabularnewline
	&	5	&	18	&	0	&	982	&	0	&	0	\tabularnewline
	&	6	&	2	&	0	&	997	&	0	&	1	\tabularnewline
\hline

CHC (corr.)
	&	1	&	1000	&	0	&	0	&	0	&	0	\tabularnewline
	&	2	&	985	&	0	&	15	&	0	&	0	\tabularnewline
	&	2.5	&	893	&	0	&	107	&	0	&	0	\tabularnewline
	&	3	&	568	&	0	&	432	&	0	&	0	\tabularnewline
	&	4	&	71	&	0	&	919	&	0	&	10	\tabularnewline
	&	5	&	8	&	0	&	887	&	0	&	105	\tabularnewline
	&	6	&	0	&	0	&	669	&	1	&	331	\tabularnewline
\hline

CHC (Eucl.)
	&	1	&	1000	&	0	&	0	&	0	&	0	\tabularnewline
	&	2	&	995	&	0	&	5	&	0	&	0	\tabularnewline
	&	2.5	&	933	&	0	&	67	&	0	&	0	\tabularnewline
	&	3	&	648	&	0	&	352	&	0	&	0	\tabularnewline
	&	4	&	139	&	0	&	861	&	0	&	0	\tabularnewline
	&	5	&	28	&	0	&	971	&	0	&	1	\tabularnewline
	&	6	&	13	&	0	&	965	&	0	&	22	\tabularnewline
\hline

CSHC
	&	1	&	1000	&	0	&	0	&	0	&	0	\tabularnewline
	&	2	&	1000	&	0	&	0	&	0	&	0	\tabularnewline
	&	2.5	&	981	&	0	&	19	&	0	&	0	\tabularnewline
	&	3	&	866	&	0	&	134	&	0	&	9	\tabularnewline
	&	4	&	328	&	0	&	672	&	0	&	1	\tabularnewline
	&	5	&	92	&	0	&	877	&	0	&	31	\tabularnewline
	&	6	&	17	&	0	&	811	&	0	&	172	\tabularnewline
\hline

  \end{tabular}}
\end{footnotesize}
\end{table}

\begin{table}[h!]
  \caption{Results of the first simulation when the values of
    $n_a$ were varied (the number of observations in the first cluster
    for the primary clusters). The clusters associated with the first
    $p_e$ rows were defined to be ``Effect 1,'' and the clusters
    associated with the final $p_e$ rows were defined to be ``Effect
    2.''  ``Effect 1/Effect 2'' means that the primary clusters
    identified by the procedure correspond to Effect 1 and the
    secondary clusters correspond to Effect 2. ``Effect 1/Neither''
    means that the primary clusters corresponded to Effect 1 but the
    secondary clusters corresponded to neither Effect 1 nor Effect
    2. PSC=preweighted sparse clustering, CHC=complementary
    hierarchical clustering (using both correlation and Euclidean
    distance), CSHC=complementary sparse hierarchical clustering.}
  \label{T:Nowak3}
{\tabcolsep=4.25pt
  \begin{tabular}{@{}cccc@{}}
   	&	n	&	Effect 1/Effect 2	&	Effect 2/Effect 1 \\ \hline
PSC	&	6	&	990	&	8	\\
					&	8	&	993	&	7	\\
					&	10	&	962	&	38	\\ \hline
CHC (corr.)	&	6	&	1000	&	0	\\
						&	8	&	1000	&	0	\\
						&	10	&
                                                1000	&	0
                                                \\ \hline
CHC (Eucl.)	&	6	&	1000	&	0	\\
						&	8	&	1000	&	0	\\
						&	10	&	1000	&	0	\\ \hline
CSHC	&	6	&	1000	&	0	\\
							&	8	&	1000	&	0	\\
							&	10	&	1000	&	0	\\ \hline
  \end{tabular}}
\end{table}

\begin{table}[h!]
\begin{footnotesize}
  \caption{Results of the first simulation when the values of $p_e$
    (the number of observations in both the primary and secondary
    clusters) were varied. The clusters associated with the first
    $p_e$ rows were defined to be ``Effect 1,'' and the clusters
    associated with the final $p_e$ rows were defined to be ``Effect
    2.'' ``Effect 1/Effect 2'' means that the primary clusters
    identified by the procedure correspond to Effect 1 and the
    secondary clusters correspond to Effect 2. ``Effect 1/Neither''
    means that the primary clusters corresponded to Effect 1 but the
    secondary clusters corresponded to neither Effect 1 nor Effect
    2. PSC=preweighted sparse clustering, CHC=complementary
    hierarchical clustering (using both correlation and Euclidean
    distance), CSHC=complementary sparse hierarchical clustering.}
  \label{T:Nowak4}
{\tabcolsep=4.25pt
  \begin{tabular}{@{}ccccccc@{}}
    &	$p_{e}$	&	Effect 1/Effect 2	&	Effect 2/Effect 1	&	Effect
1/Neither	&	Effect 2/Neither	&	Neither/Neither	\\  \hline
PSC
	&	4	&	0	&	0	&	4	&	1	&	95	\tabularnewline
	&	8	&	8	&	4	&	31	&	14	&	43	\tabularnewline
	&	12	&	33	&	18	&	31	&	16	&	2	\tabularnewline
	&	16	&	64	&	23	&	9	&	4	&	0	\tabularnewline
	&	20	&	81	&	14	&	4	&	1	&	0	\tabularnewline
	&	24	&	83	&	16	&	1	&	0	&	0	\tabularnewline
\hline

CHC (corr.)
	&	4	&	0	&	0	&	100	&	0	&	0	\tabularnewline
	&	8	&	0	&	0	&	100	&	0	&	0	\tabularnewline
	&	12	&	0	&	0	&	100	&	0	&	0	\tabularnewline
	&	16	&	79	&	0	&	21	&	0	&	0	\tabularnewline
	&	20	&	99	&	0	&	1	&	0	&	0	\tabularnewline
	&	24	&	100	&	0	&	0	&	0	&	0	\tabularnewline
\hline

CHC (Eucl.)
	&	4	&	0	&	0	&	100	&	0	&	0	\tabularnewline
	&	8	&	0	&	0	&	100	&	0	&	0	\tabularnewline
	&	12	&	0	&	0	&	100	&	0	&	0	\tabularnewline
	&	16	&	4	&	0	&	96	&	0	&	0	\tabularnewline
	&	20	&	88	&	0	&	12	&	0	&	0	\tabularnewline
	&	24	&	100	&	0	&	0	&	0	&	0	\tabularnewline
\hline

CSHC
	&	4	&	99	&	0	&	1	&	0	&	0	\tabularnewline
	&	8	&	100	&	0	&	0	&	0	&	0	\tabularnewline
	&	12	&	100	&	0	&	0	&	0	&	0	\tabularnewline
	&	16	&	100	&	0	&	0	&	0	&	0	\tabularnewline
	&	20	&	100	&	0	&	0	&	0	&	0	\tabularnewline
	&	24	&	100	&	0	&	0	&	0	&	0	\tabularnewline
\hline

  \end{tabular}}
\end{footnotesize}
\end{table}

\subsubsection{Supervised Sparse Clustering}
The mean number of misclassified observations (and associated standard
errors) when each method is applied to the final set of simulations
are shown in Table \ref{T:sim2}. Supervised sparse clustering produced
the lowest error rate of all the methods, averaging 7.2
misclassifications. Semi-supervised clustering occasionally identified
the correct clusters, but produced unsatisfactory results in many of
the simulations. Conventional sparse clustering and k-means clustering
on the principal component scores produced poor results in all the
simulated data sets. Clustering on PCA scores often produced a
singleton cluster, and semi-supervised RPMM returned a single
cluster in each simulation.

\begin{table}[h!]
  \caption{Results of the second simulation study. The following
    methods were applied to the simulated data set described in
    Section \ref{S:m_sim}: 1) supervised sparse clustering, 2) sparse
    clustering, 3) supervised clustering \citep{BT04}, 4) k-means
    clustering on the top 3 principal component (PCA) scores. The mean
    number of misclassified observations (and associated standard
    errors) are shown for each method.}
  \label{T:sim2}
{\tabcolsep=4.25pt
  \begin{tabular}{@{}ccccc@{}}

    & Sup. Sparse Clust. & Sparse Clust. & Sup. Clust. &
    Clust. on PCA \\ \hline
    Mean & 7.2  & 94.5 & 11.3 & 95.1 \\
    SE   & 0.8 & 0.1  & 0.7 & 0.1 \\ \hline
  \end{tabular}}
\end{table}

\subsection{OPPERA Data} \label{S:r_oppera}
\begin{figure}[h!]
  \caption{Feature weights for the primary, secondary and tertiary clusters
    identified by the preweighted sparse clustering method. In the
    figure below, features 1-33 are measures of experimental pain
    sensitivity, features 34-72 are measures of psychological
    distress, and features 73-116 are measures of autonomic
    function. We see that the primary clusters differ from one another
    primarily with respect to measures of autonomic functions, the
    secondary clusters differ primarily with respect to measures of
    psychological distress and the tertiary clusters differ primarily with
    respect to measures of thermal pain.}
  \centering\includegraphics[height=13cm]{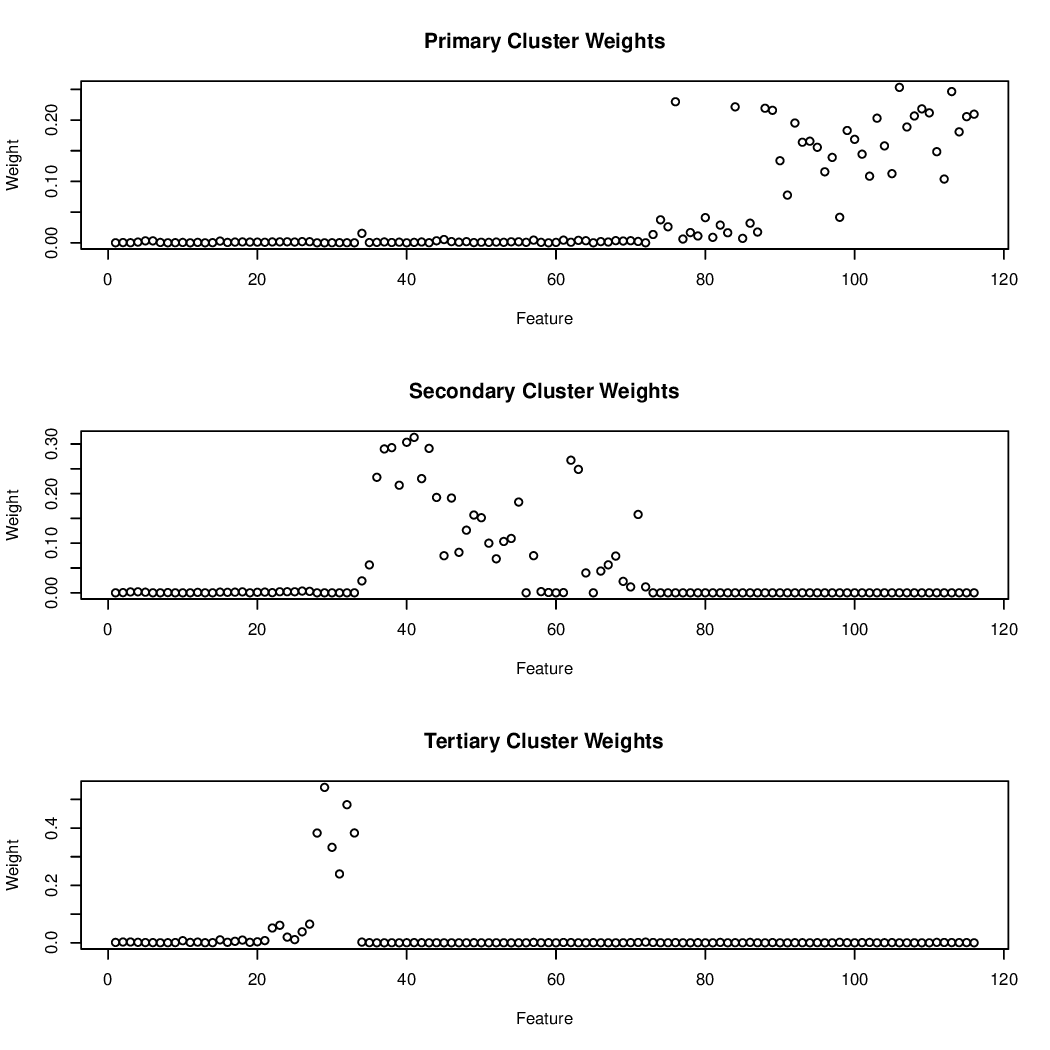}
  \label{F:psc_oppera_ws}
\end{figure}

We applied the preweighted sparse clustering method to the
OPPERA data with $k=2$. The weights for both the primary, secondary and tertiary
clusters are shown in Figure \ref{F:psc_oppera_ws}. Observe that the
measures of autonomic function had the largest feature weights for the
primary clusters, whereas the measures of psychological distress had
the largest feature weights for the secondary clusters. Measures of
thermal pain have the largest features weights for the tertiary
clusters. Thus, the preweighted sparse clustering method revealed a
biologically meaningful set of secondary and tertiary clusters that
were not identified by the conventional sparse clustering algorithm.

The associations between the primary and secondary clusters identified
by preweighted sparse clustering and hierarchical complementary
clustering are shown in Table \ref{T:psc_firstonset}. The primary
clusters identified by preweighted sparse clustering were not
significantly associated with first-onset TMD ($\text{HR}=1.2,
p=0.09$). However, the secondary clusters were associated with
first-onset TMD ($\text{HR}=1.9, p=6.5 \times 10^{-7}$). Such a result
suggests that clusters associated with an outcome of interest
(first-onset TMD in this scenario) may be obscured by a set of
clusters unrelated to the outcome of interest. The preweighted sparse
clustering method was able to identify these obscured clusters.
Neither the primary nor the secondary clusters identified by
complementary hierarchical clustering were significantly associated
with first-onset TMD.

It is interesting to note that these results are consistent with
previously published studies on the risk factors for first-onset TMD
in the OPPERA study. As observed in Figure \ref{F:psc_oppera_ws}, the
primary clusters differed mainly with respect to measures of autonomic
function whereas the secondary clusters differed mainly with respect
to measures of psychologial distress. Previous research found that the
measures of autonomic function collected in OPPERA were not associated
with first-onset TMD \citep{jG13} whereas many psychological variables
were strong predictors of first-onset TMD \citep{rF13}.

It is also interesting to compare these results with clusters
identified by \citet{eB16}. \citet{eB16} used a supervised clustering
method to identify three clusters, one of which was associated with
significantly greater risk of first-onset TMD than the other two
clusters. Participants in this high-risk cluster had higher levels of
pain sensitivity and psychological distress than participants in the
other clusters. Our current findings suggest that rather than a single
set of clusters associated with pain sensitivity and psychological
distress, there may be a primary/secondary/tertiary hierarchy of
clusters, with the secondary clusters (associated with psychological
distress) driving the association between the clusters and first-onset
TMD. Further research is needed to validate this hypothesis.

\begin{table}[h!]
  \caption{The association between the incidence of first-onset TMD
    and the primary and secondary clusters identified by preweighted
    sparse clustering method and complementary hierarchical clustering
    on the OPPERA prospective cohort data. A Cox proportional hazards
    model evaluated the null hypothesis of no association between TMD
    incidence and the cluster assignments. The hazard ratio and
    associated p-values of each cluster is reported below.}
  \label{T:psc_firstonset}
{\tabcolsep=4.25pt
  \begin{tabular}{@{}cccc@{}}
    && Hazard Ratio & P-value \\ \hline
    Preweighted Sparse Clustering & Primary Cluster & 1.2 & 0.09 \\
                                  & Secondary Cluster   & 1.9 & $6.5
                                  \times 10^{-7}$ \\  \hline
    Complementary Clustering      & Primary Cluster & 1.0 & 0.98 \\
                                  & Secondary Cluster & 1.1 & 0.32 \\
                                  \hline
  \end{tabular}}
\end{table}

Finally, we applied supervised sparse clustering (as well as several
other methods discussed earlier) to the OPPERA
data. The results are shown in Table \ref{T:ssc_oppera2}.  The two
supervised clustering methods identified clusters associated with
first-onset TMD whereas the clusters identified by sparse clustering
and clustering on the PCA scores were not associated with first-onset
TMD. This suggests that clustering methods that consider an outcome
variable may do a better job of identifying biologically relevant
clusters than methods that do not consider this information. Also,
the primary clusters identified by sparse clustering were not
associated with first-onset TMD, while the secondary clusters identified by
preweighted sparse clustering were associated with first-onset TMD.
Note that Table \ref{T:ssc_oppera2} show the results for predicted
clusters on an independent test data set, so they cannot be attributed
to overfitting.

\begin{table}[h!]
  \caption{Four different clustering methods were applied to the
    OPPERA prospective cohort training data. Each observation in the
    test data was assigned to a cluster by fitting a lasso model to
    predict the clusters on the training data and applying this model
    to the test data. The table below shows the association between each
    (predicted) cluster and first-onset TMD on the test data. For
    each method, a Cox proportional hazard model was performed to test
    the null hypothesis of no difference in survival between the two
    clusters. The hazard ratios and associated p-values are reported
    below.}
  \label{T:ssc_oppera2}
{\tabcolsep=4.25pt
  \begin{tabular}{@{}ccc@{}}
    & Hazard Ratio & P-value \\ \hline
    Supervised Sparse Clustering & 2.2 & $5.8 \times 10^{-5}$ \\
    Sparse Clustering & 1.1 & $0.69$ \\
    Supervised Clustering   & 3.1 & $3.0 \times 10^{-8}$ \\
    Clustering on PCA Scores     & 1.3 & $0.11$ \\
    Preweighted Sparse Clustering (Secondary Cluster)  & 1.8 & $6.3
    \times 10^{-4}$ \\\hline
  \end{tabular}}
\end{table}

\subsection{Leukemia Microarray Data} \label{S:r_pollack}
For each clustering method, the hazard ratio and associated
p-values for the predicted test set clusters are shown in Table
\ref{T:ssc_pollack}. Four of the methods produced clusters that were
associated with patient survival, although the clusters produced by
supervised sparse clustering were more strongly associated with
survival than the clusters produced by the other methods. (The
secondary clusters identified by preweighted sparse clustering were
not associated with survival in this case.) This indicates that
supervised sparse clustering can identify biologically meaningful and
clinically relevant clusters in high-dimensional biological data
sets. The fact that the predicted clusters were associated with
survival on an independent test set suggests that this finding is not
merely the result of overfitting.

\begin{table}[h!]
  \caption{The association between the predicted clusters for the
    test data and survival for the leukemia microarray data. For
    each method, a Cox proportional hazards model was used to test the
    null hypothesis of no difference in survival between the two
    predicted clusters. The hazard ratios and associated p-values are
    reported below.}
  \label{T:ssc_pollack}
{\tabcolsep=4.25pt
  \begin{tabular}{@{}ccc@{}}
    & Hazard Ratio & P-value \\ \hline
    Supervised Sparse Clustering & 3.4 & $6.0 \times 10^{-4}$ \\
    Sparse Clustering            & 2.2 & 0.042 \\
    Semi-Supervised Clustering   & 2.7 & 0.006 \\
    Clustering on PCA Scores     & 2.4 & 0.024 \\
    Preweighted Sparse Clustering (Secondary Cluster) & 1.9 & 0.08\\
    \hline
  \end{tabular}}
\end{table}

\section{Discussion}
\label{S:disc}
Cluster analysis is frequently used to identify subtypes in complex
data sets. In many cases, the primary objective of the cluster
analysis is to identify clusters that offer new insight into a
biological question of interest or that can be used to more precisely
phenotype (and hence diagnose and treat) a particular disease.
However, in many cases, the clusters identified by conventional
clustering methods are dominated by a subset of the features that are
not interesting biologically or clinically.

Suppose one applies a conventional clustering method and identifies
clusters that are not associated with the outcome of interest or are
not interesting biologically or clinically. One may wish to identify
secondary clusters that differ with respect to a different set of
features that may be more interesting or useful. Despite the fact that
this problem is very common in cluster analysis, relatively few
methods have been proposed to identify clusters in these
situations. As noted earlier, the idea of ``complementary clustering''
was first proposed by \citet{NT08}, and \citet{WT10} proposed an
alternative method based on sparse clustering. However, these methods
have several limitations. They can only be used with hierarchical
clustering. To our knowledge, our proposed method is the first
complementary clustering algorithm that may be applied to k-means
clustering or other clustering methods. Although we have only
considered preweighted k-means clustering in this study, our
methodology is easily applicable to sparse hierarchical clustering or
any other clustering method that can be used within the sparse
clustering framework of \citet{WT10}. Furthermore, the complementary
sparse hierarchical clustering method can be computationally
intractable when applied to data sets with numerous observations. (We
attempted to apply this method to the OPPERA data, but we were forced
to abort the procedure as it was using over 40 GB of memory.) Finally,
as observed in Sections \ref{S:r_sim} and \ref{S:r_oppera},
preweighted sparse clustering can identify clinically relevant
clusters in some situations when these existing methods do not
identify such clusters. In particular, preweighted sparse clustering
seems to perform especially well when the secondary cluster is
``difficult to detect'' (either because the mean difference between
the secondary clusters is small or the variance is large) or when
certain observations have systematically lower or higher means (and
hence are at risk of being misclassified when identifying the primary
clusters).

The problem of finding clusters that are associated with an outcome
variable has also not been studied extensively. Previously proposed
methods include the semi-supervised clustering method of \citet{BT04}
and the semi-supervised RPMM method of \citet{dK10}. Semi-supervised
clustering produces useful results in a variety of circumstances, but
the clusters produced by semi-supervised clustering can vary depending
on the choice of tuning parameters and sometimes have poor
reproducibility. Semi-supervised clustering can also fail to identify
the true clusters of interest when the association between these
clusters and the observed outcome is noisy, as we saw in Section
\ref{S:r_sim}. As noted earlier, supervised sparse clustering can
correct itself if the initial weights give zero weight to a relevant
feature or nonzero weight to an irrelevant feature (see Section
\ref{S:r_motiv}). These existing methods use a fixed set of features
that cannot be changed later, so they may produce poor results if the
features selected initially are suboptimal. Furthermore, a limitation
of semi-supervised RPMM is that it can fail to detect that clusters
exist in a data set. (Indeed, semi-supervised RPMM produced a
singleton cluster in each of the examples we considered in the present
study.) Supervised sparse clustering has been shown to overcome these
shortcomings and can produce reproducible clusters more strongly
associated with the outcome in some situations (see Section
\ref{S:r_pollack}).

It is worth noting that this general framework of selecting features
that are not associated with a primary cluster or that are associated
with an outcome variable may be applied to any clustering procedure,
not just sparse clustering. However, this approach is especially
useful in the context of sparse clustering since it tends to
``self-correct'' if the initial set of features is misspecified, as
observed in Section \ref{S:r_motiv}. This is an important advantage of
sparse clustering, since the set of initial features is unlikely to be
perfectly specified in practice.

One shortcoming of the proposed preweighted sparse clustering is the
fact that the clusters obtained may vary with respect to the choice of
the tuning parameter $s$ in the sparse clustering algorithm (see
Section \ref{S:sparse_clust}). The question of how to choose this
tuning parameter has not been studied extensively. \citet{WT10} propose
a method for choosing $s$ based on permuting the columns of the data,
but in our experience this method tends to produce values of $s$ that
are too large, which sometimes results in clusters that are not
associated (or less strongly associated) with the outcome of interest.
Choosing a smaller value of $s$ may produce better results. The
question of how to choose this tuning parameter is an area for further
study.

Despite this limitation, we believe that preweighted sparse
clustering and supervised sparse clustering are powerful tools for
solving an understudied problem. These methods can be used to identify
biologically meaningful clusters in data sets that may not be detected
by existing methods. More importantly, these methods can be used to
identify clinically relevant subtypes of diseases like TMD and cancer,
ultimately leading to better treatment options.

\section{Software}

Software in the form of R code, together with a sample
input data set and documentation is available on
request from the corresponding author (ebair@email.unc.edu). We have
plans to implement these methods in an R package.

\section{Supplementary Material}

Supplementary material (including the source code used to generate the
tables, Figures \ref{F:motiv_feature_wts} and \ref{F:psc_oppera_ws},
and the leukemia microarray data set) is available with this paper at
the journal website. The OPPERA data is available on dbGaP ({\tt
  http://www.ncbi.nlm.nih.gov/gap}).

\section*{Acknowledgments}
We wish to thank the principal investigators of the OPPERA study
(namely Richard Ohrbach, Ron Dubner, Joel Greenspan, Roger Fillingim,
Luda Diatchenko, Gary Slade, and William Maixner) for allowing us to
use the OPPERA data. This work was supported by a grant from the
National Vulvodynia Association and an NSF Graduate Research
Fellowship [to S.G.] and the National Institutes of Health [grant
numbers R03DE023592, R01HD072983, and UL1RR025747 to E.B.]. The OPPERA
study was funded by the National Institutes of Health [grant
U01DE017018].

{\it Conflict of Interest}: None declared.

\section*{References}
\bibliographystyle{elsarticle-harv}
\bibliography{bibliography.bib}

\clearpage
\setcounter{page}{1}
\renewcommand{\thesection}{S\arabic{section}}
\setcounter{section}{0}
\renewcommand{\thefigure}{S\arabic{figure}}
\setcounter{figure}{0}
\renewcommand{\thetable}{S\arabic{table}}
\setcounter{table}{0}
\renewcommand{\theequation}{S.\arabic{equation}}
\setcounter{equation}{0}

\title{Supplementary Materials for ``Identification of relevant
  subtypes via  preweighted sparse clustering'' by Sheila Gaynor and Eric Bair}\label{webappendix}
\noindent
{\bf \LARGE Web-based Supplementary Materials for ``Identification of
  relevant subtypes via  preweighted sparse clustering'' by Sheila
  Gaynor and Eric Bair}
\section{\bf Additional Simulation Results}\label{S:supp_sims}
\subsection{Mislabeled Observations in the Primary Clusters}
The objective of this simulation study was to evaluate the performance
of the complementary clustering methods when some of the observations
are assigned to the incorrect primary cluster. In other words, if
observations are assigned to the incorrect primary cluster, can the
methods still identify the correct secondary cluster? Similar to the
simulations described in Section \ref{S:m_pws_sim}, we generated a
series of 1000 $50 \times 12$ data matrices similar to
(\ref{E:Nowak_sims}) as follows:
\[
X_{ij} =
\begin{cases}
  6 + \epsilon_{ij} &\text{if $i \leq 20, j \leq 5$} \\
  -6 + \epsilon_{ij} &\text{if $i \leq 20, j > 7$} \\
  a + \epsilon_{ij} &\text{if $i \leq 20, j=6$} \\
  -a + \epsilon_{ij} &\text{if $i \leq 20, j=7$} \\
  \epsilon_{ij} &\text{if $20 < i < 30 $} \\
  3 + \epsilon_{ij} &\text{if $i \geq 30,\, j$ odd} \\
  -3 + \epsilon_{ij} &\text{if $i \geq 30,\, j$ even}
\end{cases}\text{.}
\]
The $\epsilon_{ij}$'s are iid standard normal random variables. For
$a<6$, there is a chance that observations 6 and 7 will be assigned to
the incorrect primary cluster, particularly for small values of $a$.

\begin{table}[h!]
  \begin{scriptsize}
  \caption{Results of the simulation when some observations are
    assigned may be assigned to the incorrect primary clusters. The
    clusters associated with the first 20
    rows were defined to be ``Effect 1,'' and the clusters associated
    with the final 20 rows were defined to be ``Effect 2.''
    ``Effect 1/Effect 2'' means that the primary clusters identified
    by the procedure correspond to Effect 1 and the secondary clusters
    correspond to Effect 2. ``Effect 1/Neither'' means that the
    primary clusters corresponded to Effect 1 but the secondary
    clusters corresponded to neither Effect 1 nor Effect
    2. ``Neither/Effect 2'' means that the primary clusters
    corresponded to neither Effect 1 nor Effect 2 but the secondary
    clusters corresponded to Effect 2. PSC=preweighted sparse
    clustering, CHC=complementary
    hierarchical clustering (using both correlation and Euclidean
    distance), CSHC=complementary sparse hierarchical clustering.}
  \label{T:NowakMislab}
{\tabcolsep=4.25pt
  \begin{tabular}{@{}cccccccc@{}}
    &	$a$	&	Effect 1/Effect 2	&	Effect
    2/Effect 1	&	Effect 1/Neither	&	Effect
    2/Neither & Neither/Effect 2	&	Neither/Neither	\\  \hline
PSC
	&	0.1	&	0	&	11	&0&	851	&	138	&	0	\tabularnewline
	&	0.25	&	0	&	70	&0&	834	&	96	&	0	\tabularnewline
	&	0.5	&	4	&	349	&0&	596	&	51	&	0	\tabularnewline
	&	1	&	27	&	946	&0&	24	&	3	&	0	\tabularnewline
	&	2	&	260	&	740	&0&	0	&	0	&	0	\tabularnewline
	&	4	&	946	&	54	&0&	0	&	0	&	0	\tabularnewline
	&	6	&	987	&	13	&0&	0	&	0	&	1	\tabularnewline
\hline

CHC (corr.)
	&	0.1	&	6	&	0	&	994	&0&	0	&	0	\tabularnewline
	&	0.25	&	6	&	0	&	993	&0&	0	&	1	\tabularnewline
	&	0.5	&	6	&	0	&	994	&0&	0	&	0	\tabularnewline
	&	1	&	11	&	0	&	989	&0&	0	&	0	\tabularnewline
	&	2	&	546	&	0	&	454	&0&	0	&	0	\tabularnewline
	&	4	&	1000	&	0	&	0	&0&	0	&	0	\tabularnewline
	&	6	&	1000	&	0	&	0	&0&	0	&	0	\tabularnewline
\hline

CHC (Eucl.)
	&	0.1	&	4	&	0	&	251	&0&	370	&	375	\tabularnewline
	&	0.25	&	12	&	0	&	524	&0&	225	&	239	\tabularnewline
	&	0.5	&	61	&	0	&	797	&0&	75	&	67	\tabularnewline
	&	1	&	445	&	0	&	550	&0&	0	&	0	\tabularnewline
	&	2	&	996	&	0	&	4	&0&	0	&	0	\tabularnewline
	&	4	&	1000	&	0	&	0	&0&	0	&	0	\tabularnewline
	&	6	&	1000	&	0	&	0	&0&	1	&	0	\tabularnewline
\hline

CSHC
	&	0.1	&	0	&0&	0	&	0	&	1000	&	0	\tabularnewline
	&	0.25	&	0	&0&	0	&	0	&	1000	&	0	\tabularnewline
	&	0.5	&	0	&0&	0	&	0	&	1000	&	0	\tabularnewline
	&	1	&	0	&0&	0	&	0	&	1000	&	0	\tabularnewline
	&	2	&	405	&0&	0	&	0	&	595	&	0	\tabularnewline
	&	4	&	1000	&0&	0	&	0	&	0	&	0	\tabularnewline
	&	6	&	1000	&0&	0	&	0	&	0	&	0	\tabularnewline
\hline

  \end{tabular}}
\end{scriptsize}
\end{table}

The results of this simulation study are shown in Table
\ref{T:NowakMislab}. Preweighted sparse clustering tended to choose
Effect 2 as the primary clusters for small values of $a$. However,
sometimes Effect 1 was selected as the primary cluster, and when it
was, the procedure always correctly identified Effect 2 as the
secondary cluster even if some observations in Effect 1 were assigned
to the incorrect cluster. Complementary hierarchical clustering with
correlation distance always selected Effect 1 as the primary cluster
(and nearly always identified it correctly) but always failed to
identify the secondary cluster for small values of $a$. Complementary
hierarchical clustering with Euclidean distance also always selected
Effect 1 as the primary cluster, although not always
correctly. Interestingly, it usually failed to identify Effect 2 when
Effect 1 was identified correctly, but it correctly identified Effect
2 about half the time when Effect 1 was identified
incorrectly. Complementary sparse hierarchical clustering always
selected Effect 1 as the primary cluster. It never identified Effect 1
correctly for smaller values of $a$. In each simulation, however, it
correctly identified Effect 2 as the secondary cluster. Thus, we see
that preweighted sparse clustering can identify secondary clusters
even if some observations are assigned to the incorrect primary
cluster.

\subsection{Correlated Features}
The objective of this simulation study was to evaluate the performance
of the complementary clustering methods when the data was
correlated. We generated a series of $50 \times 12$ data matrices
according to (\ref{E:Nowak_sims}). However, rather than being iid, the
$\epsilon_{ij}$'s were multivariate normal with mean 0 with covariance
matrix $\Sigma$ with $\sigma_{i,j}=\rho^{|i-j|/5}$, where
$\sigma_{i,j}$ denotes the $(i,j)$th element of $\Sigma$.

The results of this simulation study are shown in Table
\ref{T:NowakCorr}. Complementary hierarchical clustering using
Euclidean distance occasionally failed to identify Effect 2 when
$\rho$ (and hence the correlation between adjacent features) was high,
although it failed less than 10\% of the time. The remaining methods
failed to identify either effect less than 1\% of the time suggesting
that they are robust to correlated features.

\begin{table}[h!]
  \begin{small}
  \caption{Results of the simulation when the features were
    correlated. The clusters associated with the first 20
    rows were defined to be ``Effect 1,'' and the clusters associated
    with the final 20 rows were defined to be ``Effect 2.''
    ``Effect 1/Effect 2'' means that the primary clusters identified
    by the procedure correspond to Effect 1 and the secondary clusters
    correspond to Effect 2. ``Effect 1/Neither'' means that the
    primary clusters corresponded to Effect 1 but the secondary
    clusters corresponded to neither Effect 1 nor Effect
    2. PSC=preweighted sparse clustering, CHC=complementary
    hierarchical clustering (using both correlation and Euclidean
    distance), CSHC=complementary sparse hierarchical clustering.}
  \label{T:NowakCorr}
{\tabcolsep=4.25pt
  \begin{tabular}{@{}ccccccc@{}}
    &	$\rho$	&	Effect 1/Effect 2	&	Effect
    2/Effect 1	&	Effect 1/Neither	&  Effect 2/Neither
    &	Neither/Neither	\\  \hline
PSC
	&	0.1	&	957	&	43	&	0	&	0	&	0	\tabularnewline
	&	0.2	&	933	&	66	&	1	&	0	&	0	\tabularnewline
	&	0.3	&	917	&	80	&	2	&	1	&	0	\tabularnewline
	&	0.4	&	893	&	104	&	2	&	0	&	1	\tabularnewline
	&	0.5	&	894	&	100	&	4	&	2	&	0	\tabularnewline
\hline

CHC (corr.)
	&	0.1	&	1000	&	0	&	0	&	0	&	0	\tabularnewline
	&	0.2	&	1000	&	0	&	0	&	0	&	1	\tabularnewline
	&	0.3	&	998	&	0	&	2	&	0	&	0	\tabularnewline
	&	0.4	&	998	&	0	&	2	&	0	&	0	\tabularnewline
	&	0.5	&	994	&	0	&	2	&	0	&	0	\tabularnewline
\hline

CHC (Eucl.)
	&	0.1	&	1000	&	0	&	0	&	0	&	0	\tabularnewline
	&	0.2	&	997	&	0	&	3	&	0	&	0	\tabularnewline
	&	0.3	&	985	&	0	&	15	&	0	&	0	\tabularnewline
	&	0.4	&	952	&	0	&	48	&	0	&	0	\tabularnewline
	&	0.5	&	925	&	0	&	75	&	0	&	0	\tabularnewline
\hline

CSHC
	&	0.1	&	1000	&	0	&	0	&	0	&	0	\tabularnewline
	&	0.2	&	1000	&	0	&	0	&	0	&	0	\tabularnewline
	&	0.3	&	1000	&	0	&	0	&	0	&	0	\tabularnewline
	&	0.4	&	1000	&	0	&	0	&	0	&	0	\tabularnewline
	&	0.5	&	999	&	0	&	1	&	0	&	0	\tabularnewline
\hline

  \end{tabular}}
\end{small}
\end{table}

\subsection{Three Primary Clusters}
The objective of these simulations studies were to evaluate the
performance of the complementary clustering methods when there were
three (rather than two) primary clusters. For the first study,
we modified (\ref{E:Nowak_sims}) to generate a series of 1000 $50
\times 12$ data matrices as follows:
\[
X_{ij} =
\begin{cases}
  6 + \epsilon_{ij} &\text{if $i \leq 20, j \leq 4$} \\
  -6 + \epsilon_{ij} &\text{if $i \leq 20, j > 8$} \\
  \epsilon_{ij} &\text{if $20 < i < 31 $} \\
  b + \epsilon_{ij} &\text{if $i \geq 31,\, j$ odd} \\
  -b + \epsilon_{ij} &\text{if $i \geq 31,\, j$ even}
\end{cases}\text{.}
\]
Note that features 1-20 have mean 0 for features 5-8, producing a
third primary cluster. We applied each complementary hierarchical
clustering method first with $k=3$ to identify the primary clusters
and then with $k=2$ to identify the secondary clusters.

The results of this simulation study are shown in Table
\ref{T:NowakThree}. Preweighted sparse clustering occasionally
identifies both Effect 1 and Effect 2 correctly (particularly when $b$
is small). However, it often fails to identify Effect 2 for small $b$
and fails to identify Effect 1 for large $b$, and occasionally it
fails to detect both effects. Complementary hierarchical clustering
with correlation distance has a disastrous performance in this
scenario, failing to detect both effects in essentially all simulation
scenarios. When Euclidean distance is used, it detects Effect 2 but
not Effect 1 for small values of $b$. Its performance is excellent
when $b=2$, detecting both effects 97.5\% of the time. Oddly, when
$b=3$, it detects both effects 27.8\% of the time but fails to detect
either effect 71.3\% of the time. Complementary sparse hierarchical
clustering always detects Effect 2 and also always detect Effect 1
when $b=2$ or $b=3$.

\begin{table}[h!]
  \begin{small}
  \caption{Results of the simulation with three primary clusters and a
    means of $\pm b$ in the secondary clusters. The
    clusters associated with the first 20
    rows were defined to be ``Effect 1,'' and the clusters associated
    with the final 20 rows were defined to be ``Effect 2.''
    ``Effect 1/Effect 2'' means that the primary clusters identified
    by the procedure correspond to Effect 1 and the secondary clusters
    correspond to Effect 2. ``Effect 1/neither'' means that the
    primary clusters corresponded to Effect 1 but the secondary
    clusters corresponded to neither Effect 1 nor Effect
    2. ``Neither/Effect 2'' means that the primary clusters
    corresponded to neither Effect 1 nor Effect 2 but the secondary
    clusters corresponded to Effect 2. PSC=preweighted sparse
    clustering, CHC=complementary
    hierarchical clustering (using both correlation and Euclidean
    distance), CSHC=complementary sparse hierarchical clustering.}
  \label{T:NowakThree}
{\tabcolsep=4.25pt
  \begin{tabular}{@{}cccccc@{}}
    &	$b$	&	Effect 1/Effect 2	&	Effect
    1/Neither	&  Neither/Effect 2 &	Neither/Neither	\\  \hline
PSC
	&	0.5	&	336	&	639	&	0	&	25	\tabularnewline
	&	0.75	&	532	&	30	&	1	&	437	\tabularnewline
	&	1	&	223	&	1	&	0	&	776	\tabularnewline
	&	2	&	12	&	0	&	942	&	46	\tabularnewline
	&	3	&	2	&	0	&	988	&	10	\tabularnewline
\hline

CHC (corr.)
	&	0.5	&	0	&	1	&	0	&	999	\tabularnewline
	&	0.75	&	0	&	1	&	0	&	999	\tabularnewline
	&	1	&	0	&	0	&	0	&	1000	\tabularnewline
	&	2	&	0	&	0	&	0	&	1000	\tabularnewline
	&	3	&	0	&	0	&	0	&	1000	\tabularnewline
\hline

CHC (Eucl.)
	&	0.5	&	0	&	1000	&	0	&	0	\tabularnewline
	&	0.75	&	0	&	1000	&	0	&	0	\tabularnewline
	&	1	&	0	&	1000	&	0	&	0	\tabularnewline
	&	2	&	975	&	25	&	0	&	0	\tabularnewline
	&	3	&	278	&	0	&	9	&	713	\tabularnewline
\hline

CSHC
	&	0.5	&	0	&	1000	&	0	&	0	\tabularnewline
	&	0.75	&	0	&	1000	&	0	&	0	\tabularnewline
	&	1	&	1	&	999	&	0	&	0	\tabularnewline
	&	2	&	1000	&	0	&	0	&	0	\tabularnewline
	&	3	&	1000	&	0	&	0	&	0	\tabularnewline
\hline

  \end{tabular}}
\end{small}
\end{table}
We also considered a second simulation with three primary clusters
where we varied the number of features that defined the primary
clusters. Specifically, we modified (\ref{E:Nowak_sims}) to generate
1000 data matrices with $(p_b+120) \times 12$ observations as follows:
\[
X_{ij} =
\begin{cases}
  6 + \epsilon_{ij} &\text{if $i \leq p_b, j \leq 4$} \\
  -6 + \epsilon_{ij} &\text{if $i \leq p_b, j > 8$} \\
  \epsilon_{ij} &\text{if $p_b < i < p_b+101 $} \\
  b + \epsilon_{ij} &\text{if $i \geq p_b+101,\, j$ odd} \\
  -b + \epsilon_{ij} &\text{if $i \geq p_b+101,\, j$ even}
\end{cases}\text{.}
\]
In other words, $p_b$ features define Effect 1, 20 features define
Effect 2, and 100 features are associated with neither Effect 1 nor
Effect 2.

\begin{table}[h!]
  \begin{small}
  \caption{Results of the simulation with three primary clusters and
    $p_b$ clusters defining the primary clusters. The
    clusters associated with the first $p_b$
    rows were defined to be ``Effect 1,'' and the clusters associated
    with the final 20 rows were defined to be ``Effect 2.''
    ``Effect 1/Effect 2'' means that the primary clusters identified
    by the procedure correspond to Effect 1 and the secondary clusters
    correspond to Effect 2. ``Effect 1/Neither'' means that the
    primary clusters corresponded to Effect 1 but the secondary
    clusters corresponded to neither Effect 1 nor Effect
    2. ``Neither/Effect 2'' means that the primary clusters
    corresponded to neither Effect 1 nor Effect 2 but the secondary
    clusters corresponded to Effect 2. PSC=preweighted sparse
    clustering, CHC=complementary
    hierarchical clustering (using both correlation and Euclidean
    distance), CSHC=complementary sparse hierarchical clustering.}
  \label{T:NowakThree2}
{\tabcolsep=4.25pt
  \begin{tabular}{@{}cccccc@{}}
    &	$p_b$	&	Effect 1/Effect 2	&	Effect
    1/Neither	&  Neither/Effect 2 &	Neither/Neither	\\  \hline
PSC
	&	20	&	4	&	0	&	690	&	306	\tabularnewline
	&	30	&	2	&	0	&	1	&	997	\tabularnewline
	&	50	&	266	&	4	&	0	&	730	\tabularnewline
	&	100	&	1000	&	0	&	0	&	0	\tabularnewline
	&	200	&	997	&	3	&	0	&	0	\tabularnewline
\hline

CHC (corr.)
	&	20	&	0	&	0	&	926	&	74	\tabularnewline
	&	30	&	0	&	0	&	951	&	49	\tabularnewline
	&	50	&	0	&	0	&	826	&	174	\tabularnewline
	&	100	&	0	&	0	&	169	&	831	\tabularnewline
	&	200	&	0	&	0	&	0	&	1000	\tabularnewline
\hline

CHC (Eucl.)
	&	20	&	270	&	0	&	4	&	726	\tabularnewline
	&	30	&	1000	&	0	&	0	&	0	\tabularnewline
	&	50	&	967	&	33	&	0	&	0	\tabularnewline
	&	100	&	0	&	1000	&	0	&	0	\tabularnewline
	&	200	&	0	&	1000	&	9	&	0	\tabularnewline
\hline

CSHC
	&	20	&	1000	&	0	&	0	&	0	\tabularnewline
	&	30	&	1000	&	0	&	0	&	0	\tabularnewline
	&	50	&	1000	&	0	&	0	&	0	\tabularnewline
	&	100	&	1000	&	0	&	0	&	0	\tabularnewline
	&	200	&	1000	&	0	&	0	&	0	\tabularnewline
\hline

  \end{tabular}}
\end{small}
\end{table}
The results of this simulation scenario are shown in Table
\ref{T:NowakThree2}. Complementary sparse hierarchical clustering
performed extremely well in this scenario, correctly identifying both
effects in every simulation. Complementary hierarchical clustering
with correlation distance performed poorly. It never correctly
identified Effect 1, and often failed to detect both effects for
larger values of $p_b$. Preweighted sparse clustering tended to
perform poorly for small values of $p_b$, but it correctly identified
both effects for larger values of $p_b$. In contrast, complementary
hierarchical clustering with Euclidean distance tended to perform
poorly for larger values of $p_b$ but generally produced good results
for smaller values of $p_b$ (although it frequently failed to detect
both effects when $p_b=20$).

\subsection{Features Associated with Both Primary and Secondary
  Clusters}
The objective of this simulation study was to evaluate the performance
of the complementary clustering methods when there was overlap between
the set of features that defined the primary and secondary
clusters. First, a series of 1000 $50 \times 100$ data sets were
generated using a modified form of (\ref{E:motiv}) in Section
\ref{S:motiv}:
\[
X_{ij} =
\begin{cases}
  6 + \epsilon_{ij} &\text{if $i \leq 10, j \leq 50$} \\
  -6 + \epsilon_{ij} &\text{if $i \leq 10, j > 50$} \\
  6+b + \epsilon_{ij} &\text{if $11 \leq i \leq 20$, $j \leq 50$, and
    $j$ odd} \\
  6-b + \epsilon_{ij} &\text{if $11 \leq i \leq 20$, $j \leq 50$, and
    $j$ even} \\
  -6+b + \epsilon_{ij} &\text{if $11 \leq i \leq 20$, $j \leq 50$, and
    $j$ odd} \\
  -6-b + \epsilon_{ij} &\text{if $11 \leq i \leq 20$, $j \leq 50$, and
    $j$ even} \\
  3 + \epsilon_{ij} &\text{if $i \leq 21 \leq 30$, $j$ odd} \\
  -3 + \epsilon_{ij} &\text{if $i \leq 21 \leq 30$, $j$ even} \\
  \epsilon_{ij} &\text{if $i > 30$}
\end{cases}\text{.}
\]
Once again, the $\epsilon_{ij}'s$ are iid standard normal random
variables.

The results of this simulation scenario are shown in Table
\ref{T:NowakOverlap}. Preweighted sparse clustering always identified
Effect 1 correctly. The likelihood of identifying Effect 2 increased as
$b$ increased (and Effect 2 was always identified correctly for $b
\geq 2$). The results were comparable for complementary hierarchical
clustering with correlation distance, although it was less likely to
identify Effect 2 for small values of $b$. The results for
complementary hierarchical clustering with Euclidean distance were
generally comparable to the results for preweighted sparse clustering,
although it frequently failed to identify both effects when
$b=6$. Complementary sparse hierarchical clustering performed
poorly. It never identified Effect 2, and it also never identified
Effect 1 when $b=6$.

\begin{table}[h!]
  \begin{small}
  \caption{Results of the simulation with overlapping features
    defining the clusters. The clusters associated with the first 20
    rows were defined to be ``Effect 1,'' and the clusters associated
    with rows 11-30 were defined to be ``Effect 2.''
    ``Effect 1/Effect 2'' means that the primary clusters identified
    by the procedure correspond to Effect 1 and the secondary clusters
    correspond to Effect 2. ``Effect 1/neither'' means that the
    primary clusters corresponded to Effect 1 but the secondary
    clusters corresponded to neither Effect 1 nor Effect
    2. PSC=preweighted sparse clustering, CHC=complementary
    hierarchical clustering (using both correlation and Euclidean
    distance), CSHC=complementary sparse hierarchical clustering.}
  \label{T:NowakOverlap}
{\tabcolsep=4.25pt
  \begin{tabular}{@{}cccccc@{}}
    &	$b$	&	Effect 1/Effect 2	&	Effect
    2/Effect 1	&  Effect 1/Neither &	Neither/Neither	\\  \hline
PSC
	&	0.5	&	1	&	0	&	999	&	0	\tabularnewline
	&	0.75	&	314	&	0	&	686	&	0	\tabularnewline
	&	1	&	910	&	0	&	90	&	0	\tabularnewline
	&	2	&	1000	&	0	&	0	&	0	\tabularnewline
	&	3	&	1000	&	0	&	0	&	0	\tabularnewline
	&	6	&	496	&	504	&	0	&	0	\tabularnewline
\hline

CHC (corr.)
	&	0.5	&	0	&	0	&	1000	&	0	\tabularnewline
	&	0.75	&	24	&	0	&	976	&	0	\tabularnewline
	&	1	&	514	&	0	&	486	&	0	\tabularnewline
	&	2	&	1000	&	0	&	0	&	0	\tabularnewline
	&	3	&	1000	&	0	&	0	&	0	\tabularnewline
	&	6	&	478	&	522	&	0	&	0	\tabularnewline
\hline

CHC (Eucl.)
	&	0.5	&	0	&	0	&	1000	&	0	\tabularnewline
	&	0.75	&	325	&	0	&	675	&	0	\tabularnewline
	&	1	&	900	&	0	&	100	&	0	\tabularnewline
	&	2	&	1000	&	0	&	0	&	0	\tabularnewline
	&	3	&	1000	&	0	&	0	&	0	\tabularnewline
	&	6	&	275	&	243	&	0	&	482	\tabularnewline
\hline

CSHC
	&	0.5	&	0	&	0	&	1000	&	0	\tabularnewline
	&	0.75	&	0	&	0	&	1000	&	0	\tabularnewline
	&	1	&	0	&	0	&	1000	&	0	\tabularnewline
	&	2	&	0	&	0	&	1000	&	0	\tabularnewline
	&	3	&	0	&	0	&	939	&	61	\tabularnewline
	&	6	&	0	&	0	&	0	&	1000	\tabularnewline
\hline

  \end{tabular}}
\end{small}
\end{table}
A second set of 1000 simulated $50 \times 100$ data sets was generated
using the following procedure:
\begin{itemize}
  \item Set $X_{ij}=\epsilon_{ij}$, where the $\epsilon_{ij}$'s are
    iid standard normal random variables.
  \item If $i \leq 20$ and $j \leq 50-n_a$, let $X_{ij}=X_{ij}+6$.
  \item If $i \leq 20$ and $j > 50-n_a$, let $X_{ij}=X_{ij}-6$.
  \item If $11 \leq i \leq 30$ and $26 \leq j \leq 75$, let
    $X_{ij}=X_{ij}+3$.
  \item If $11 \leq i \leq 30$ and $j \leq 25$ or $j>76$, let
    $X_{ij}=X_{ij}-3$.
\end{itemize}
In other words, the primary cluster (Effect 1) consists of $50-n_a$
observations with mean $6$ and $50+n_a$ observations with mean
$-6$. The secondary cluster (Effect 2) consists of 50 observations
with mean $3$ and 50 observations with mean mean $-3$. These effects
are also additive, so the mean of a feature belonging to both the
primary and secondary clusters has a mean equal to the sum of the
means of the individual effects. Moreover, for $n_a>0$, membership in
the primary and secondary clusters are not independent: observations
with a mean of 6 in the primary cluster for features 1-20 are more
likely to belong to have a mean of 3 in the secondary cluster for
features 11-30.

\begin{table}[h!]
  \begin{small}
  \caption{Results of the simulation with overlapping features
    defining the clusters. The clusters associated with the first 20
    rows were defined to be ``Effect 1,'' and the clusters associated
    with rows 11-30 were defined to be ``Effect 2.''
    ``Effect 1/Effect 2'' means that the primary clusters identified
    by the procedure correspond to Effect 1 and the secondary clusters
    correspond to Effect 2. ``Effect 1/neither'' means that the
    primary clusters corresponded to Effect 1 but the secondary
    clusters corresponded to neither Effect 1 nor Effect
    2. PSC=preweighted sparse clustering, CHC=complementary
    hierarchical clustering (using both correlation and Euclidean
    distance), CSHC=complementary sparse hierarchical clustering.}
  \label{T:NowakOverlap2}
{\tabcolsep=4.25pt
  \begin{tabular}{@{}ccccc@{}}
    &	$n_a$	&	Effect 1/Effect 2	&	Effect
    1/Neither	&  Neither/Neither	\\  \hline
PSC
	&	0	&	1000	&	0	&	0	\tabularnewline
	&	5	&	1000	&	0	&	0	\tabularnewline
	&	10	&	1000	&	0	&	0	\tabularnewline
	&	15	&	1000	&	0	&	0	\tabularnewline
	&	20	&	139	&	861	&	0	\tabularnewline
	&	25	&	154	&	846	&	0	\tabularnewline
\hline

CHC (corr.)
	&	0	&	1000	&	0	&	0	\tabularnewline
	&	5	&	1000	&	0	&	0	\tabularnewline
	&	10	&	1000	&	0	&	0	\tabularnewline
	&	15	&	998	&	2	&	0	\tabularnewline
	&	20	&	935	&	65	&	0	\tabularnewline
	&	25	&	3	&	997	&	0	\tabularnewline
\hline

CHC (Eucl.)
	&	0	&	1000	&	0	&	0	\tabularnewline
	&	5	&	1000	&	0	&	0	\tabularnewline
	&	10	&	1000	&	0	&	0	\tabularnewline
	&	15	&	1000	&	0	&	0	\tabularnewline
	&	20	&	985	&	0	&	15	\tabularnewline
	&	25	&	0	&	0	&	1000	\tabularnewline
\hline

CSHC
	&	0	&	0	&	914	&	86	\tabularnewline
	&	5	&	0	&	948	&	52	\tabularnewline
	&	10	&	0	&	963	&	37	\tabularnewline
	&	15	&	0	&	988	&	12	\tabularnewline
	&	20	&	0	&	1000	&	0	\tabularnewline
	&	25	&	0	&	1000	&	0	\tabularnewline
\hline

  \end{tabular}}
\end{small}
\end{table}
The results of this simulation scenario are shown in Table
\ref{T:NowakOverlap2}. Both preweighted sparse clustering and
complementary hierarchical clustering sometimes failed to identify the
secondary clusters for large values of $n_a$. Complementary
hierarchical clustering produced better results when $n_a=20$,
although preweighted sparse clustering produced better results when
$n_a=25$. Complementary sparse hierarchical clustering produced poor
results for all values of $n_a$. It never identified Effect 2
correctly, and it sometimes also failed to identify Effect 1 for
smaller values of $n_a$.
\label{lastpage}

\end{document}